\documentclass[letterpaper,12pt]{article}
\usepackage{jheppub} %
\usepackage{caption}
\usepackage{subcaption}
\usepackage{wrapfig}
\usepackage{color,soul}
\usepackage{lscape}
\usepackage{rotating}
\usepackage{footmisc}
\usepackage{array}
\usepackage{csquotes}
\usepackage{lmodern}
\usepackage{lastpage}
\renewcommand{\scshape}{} %

\graphicspath{{Images/}}

\newcommand{\Figure}[1]{\text{Figure~}\ref{#1}}

\renewcommand{\thefootnote}{\fnsymbol{footnote}}

\newcommand{\authorline}[1]{\noindent \textbf{\small #1}}

\RequirePackage{fancyhdr}

\def\stryear{2025}
\def\strid{006}
\def\strnum{SNOLAB-STR-\stryear-\strid}
\makeatletter\def\@fpheader{\strnum}\makeatother

\title{\boldmath Community Report from the 2025 SNOLAB Future Projects Workshop}

\abstract{SNOLAB hosts a biannual Future Projects Workshop (FPW) with the goal of encouraging future project stakeholders to present ideas, concepts, and needs for experiments or programs that could one day be hosted at SNOLAB. The 2025 FPW was held in the larger context of a 15-year planning exercise requested by the Canada Foundation for Innovation. This report collects input from the community, including both contributions to the workshop and contributions that could not be scheduled in the workshop but nonetheless are important to the community.}

\author[1]{M. Diamond* \note{Corresponding author.}} %
\emailAdd{m.diamond@mail.utoronto.ca}

\author[2]{P.~Abbamonte,} 
\author[3]{A.~Arvanitaki,} 
\author[4]{D.~M.~Asner,} 
\author[2]{D.~Balut,} 
\author[5]{D.~Baxter,} 
\author[6]{C.~Blanco,} 
\author[7]{D.~Boreham,} 
\author[8]{M.~Boulay,} 
\author[9]{B.~Broerman,} 
\author[10]{T.~Brunner,} 
\author[4,10,11]{E.~Caden,} 
\author[12]{A.~Chavarria,}
\author[9]{M.~Chen,} 
\author[13]{J.~P.~Davis,} 
\author[5,14]{A.~Drlica-Wagner,} 
\author[5]{J.~Estrada,}
\author[4]{N. Fatemighomi,}
\author[5]{J.~Foster,}
\author[15]{D.~Freedman,} 
\author[16]{C.~Gao,} 
\author[4,11]{J.~Hall,}
\author[4]{S.~Hall,}
\author[17]{W.~Halperin,} 
\author[13]{M. Hirschel,} 
\author[18]{N.~Hoch,} 
\author[1]{Z. Hong,} 
\author[19,20]{A.~Ianni,} 
\author[4,11]{C.~Jillings,} 
\author[15]{D.~Johnson,} 
\author[1]{Y.~Kahn,} 
\author[13]{C.~B.~Krauss,} 
\author[11]{T.~Laframboise,} 
\author[9,21]{M. Lai,} 
\author[7,11]{M.~R.~Lapointe,} 
\author[6]{B.~Lillard,} 
\author[22]{W.H.~Lippincott,}
\author[23]{H.~Ma,} 
\author[14]{E.~Marrufo Villalpando,} 
\author[24]{K. Mistry,} 
\author[17]{M.~Nguyen,} 
\author[2]{J.~Oh,} 
\author[25]{A.~Radick,} 
\author[11]{H.~Reaume,} 
\author[14]{B.~Roach,} 
\author[26]{J.~Sch\"{u}tte-Engel,} 
\author[27]{S.~Scorza,} 
\author[17]{J.~W.~Scott,} 
\author[4,9,11]{S.~J.~Sekula,} 
\author[8]{D.~Sinclair,} 
\author[7,11]{C.~Thome,} 
\author[2]{L.~Thompson,} 
\author[5]{J.~Tiffenberg,} 
\author[18]{K.~J.~Vetter,} 
\author[14]{A.~Williams,} 
\author[18]{L.~A. Winslow,} 
\author[28]{M.~Wurm}

\affiliation[1]{Department of Physics, University of Toronto, Toronto, ON, M5S 1A7, Canada}
\affiliation[2]{Department of Physics, University of Illinois Urbana-Champaign, Urbana, IL, 61801, USA}
\affiliation[3]{Perimeter Institute for Theoretical Physics, Waterloo, Ontario, N2L 2Y5, Canada}
\affiliation[4]{SNOLAB, Lively, ON, P3Y 1N2, Canada}
\affiliation[5]{Fermi National Accelerator Laboratory, Batavia, IL 60510, USA}
\affiliation[6]{Department of Physics, Penn State University, State College, PA, 16802, USA}
\affiliation[7]{Medical Sciences Division, NOSM University, Sudbury, ON, P3E 2C6, Canada}
\affiliation[8]{Department of Physics, Carleton University, Ottawa, ON, K1S 5B6, Canada}
\affiliation[9]{Department of Physics, Engineering Physics, and Astronomy, Queen's University, Kingston, ON, K7L 3N6, Canada}
\affiliation[10]{Department of Physics, McGill University, Montréal, QC, H3A 2T8, Canada}
\affiliation[11]{School of Natural Sciences, Laurentian University, Sudbury, ON, P3E 2C6, Canada}
\affiliation[12]{Center for Experimental Nuclear Physics and Astrophysics, University of Washington, Seattle, WA, 98195, USA}
\affiliation[13]{Department of Physics, University of Alberta, Edmonton, AB T6G 2E9, Canada}
\affiliation[14]{Kavli Institute of Cosmological Physics, University of Chicago, Chicago, IL, 60637, USA}
\affiliation[15]{Department of Chemistry, Massachusetts Institute of Technology, Cambridge, MA, 02139, USA}
\affiliation[16]{Department of Physics, Southern University of Science and Technology, Shenzhen, China}
\affiliation[17]{Department of Physics and Astronomy, Northwestern University, Evanston, IL, 60208, USA}
\affiliation[18]{Laboratory for Nuclear Science, Massachusetts Institute of Technology, Cambridge, MA, 02139, USA}
\affiliation[19]{Physics Department, Princeton University, New Jersey, 08544, USA}
\affiliation[20]{INFN Laboratori Nazionali del Gran Sasso, Assergi (AQ) 67100, Italy}
\affiliation[21]{Department of Physics \& Astronomy, University of California, Riverside, Riverside, CA, 92521, USA}
\affiliation[22]{Department of Physics, University of California, Santa Barbara,  Santa Barbara, CA 93106-9530, USA}
\affiliation[23]{Department of Engineering Physics, Tsinghua University, Beijing, 100084, China}
\affiliation[24]{Department of Physics, University of Texas at Arlington, Arlington, TX, 76019, USA}
\affiliation[25]{Department of Physics, University of Oregon, Eugene, OR, 97403, USA}
\affiliation[26]{Department of Physics, University of California, Berkeley, Berkeley, CA, 94720, USA}
\affiliation[27]{Universit\'e Grenoble Alpes, CNRS, Grenoble INP, LPSC-IN2P3, Grenoble, 38000, France}
\affiliation[28]{Institut für Physik, Johannes Gutenberg-Universit\:at Mainz, Mainz, 55128, Germany}

\begin{document} 
\maketitle
\flushbottom

\pagestyle{fancy}
\fancyhead{} %
\fancyfoot{} %
\fancyfoot[R]{Page \thepage\ of \pageref{LastPage}}
\fancyfoot[L]{Summer 2025\\\strnum}

\setcounter{footnote}{0} %
\renewcommand{\thefootnote}{\arabic{footnote}}

\newpage

\section{A Summary of the 2025 SNOLAB Future Projects Workshop for the General Public}
\label{sec:summary_public}

\authorline{Co-Authors: N.~Fatemighomi, S.~Hall, S.~J.~Sekula}

\bigskip

SNOLAB is a world-class scientific facility located 2~km underground in an active nickel mine in Lively, ON, Canada. SNOLAB consists of two major components. The main underground laboratory hosts 5000~$\mathrm{m^2}$ of science-grade cleanroom space, making it the simultaneously deepest and cleanest laboratory of its kind. On the surface is a three-storey office and laboratory building. The underground lab is so deep to protect hosted experiments from the naturally occurring space radiation that rains down on the surface of the Earth. SNOLAB has the lowest such radiation level of existing laboratories.

The laboratory is operated by a diverse and talented 140-person staff whose skills range across many disciplines from cleanliness to multiple technical fields and trades, to engineering and design, project management, scientific research, safety, information technology, communications, education, and outreach. The laboratory hosts and supports projects from an international community of 1,200 scientists, technicians, and engineers. This community includes many early career personnel in the student or post-graduate professional stages of their training experience. SNOLAB hosts a range of scientific experiments ranging from astronomy and astrophysics to biology, chemistry, health sciences, environmental monitoring, and geoscience. The laboratory is a catalyst for national and international collaboration in these fields. KPMG concluded in 2022 that SNOLAB generates \$3.42 in economic activity for every \$1 invested in the laboratory from any funding source~\cite{SNOLAB_annual_report}.

Every two years, the laboratory hosts a "Future Projects Workshop". This meeting invites the community to reflect on opportunities that offer promising directions for the laboratory. This is also a chance for the community to engage in dialogue, develop new ideas, and forge new partnerships. The 2025 meeting synergistically overlapped with an effort to develop a 15-year laboratory plan and vision. That plan was requested by SNOLAB's primary funding agency, the Canada Foundation for Innovation, in 2024 with a delivery date of September, 2025.

Workshop participants came from many disciplines and from places across the globe. They spoke on a range of subjects. For example, there was a discussion of the biological effects of low-radiation environments on living organisms and opportunities to lead and coordinate work across multiple internationally sited underground laboratories. There were conversations and advocacy around the need to build future experiments for astrophysics. Some of these rely on small, ultra-sensitive technologies whose full promise has not yet been recognized. Others rely on time-tested technologies that now require scaling up to such large sizes they would fill, or exceed, the current volume of any existing cavern in the underground facility.

Participants discussed novel uses for quantum sensing technology, including the detection of gravitational waves from distant cataclysmic events. The community drove conversations between currently distinct collaborations about the idea of combining the greatest features of their technologies into advanced, multidisciplinary experiments with potentially ground-breaking science reach. The spirit of scientific pursuit for answers to unanswered questions, combined with the recognition of the scientific and engineering training opportunities these afford, was strong at this meeting.

The Future Projects Workshop was also one of the rare opportunities for peer underground laboratories to share their own plans and visions for both lab-specific portfolios and cross-laboratory collaboration, including collaborative projects, personnel exchange, and information sharing. Underground laboratory representatives provided this information openly at the workshop. This was also a chance for the community to think more about research funding sources and provided an opportunity to reflect on ways to diversify that funding.

The participants were highly motivated to share their ideas in this open and engaging forum. SNOLAB was recognized for its ability to develop, manage, and deploy projects that have never been done before. The lab leads on solving difficult problems in a challenging underground environment. Participants were also drawn to the role that the laboratory plays in sharing information about discoveries, ideas, and milestones with the global community.

The community participants, through the perspective of their projects and ideas, identified a number of areas where SNOLAB could enhance its capabilities. These span a wide range. There was advocacy for more overall laboratory space as well as larger individual spaces for hosting even more massive projects underground. There was a recognition of a need for more widely accessible space on the surface.

Participants pointed to needs for new capabilities to meet future challenges, like expertise in high-pressure gas or ultra-cold environments. A strong emphasis from participants was on growing the capabilities of an already-low-radiation environment, making the lab even more sensitive to small effects and even better at protecting against those that remain. One key request was for more small facilities where it is possible to test new ideas and technologies. 

The community overall is interested in a healthy portfolio of smaller-scale, high-impact experiments and long-timescale, large-scale projects.

\newpage
\section{Introduction}

\begin{figure}[ht]
    \centering
    \includegraphics[width=\linewidth]{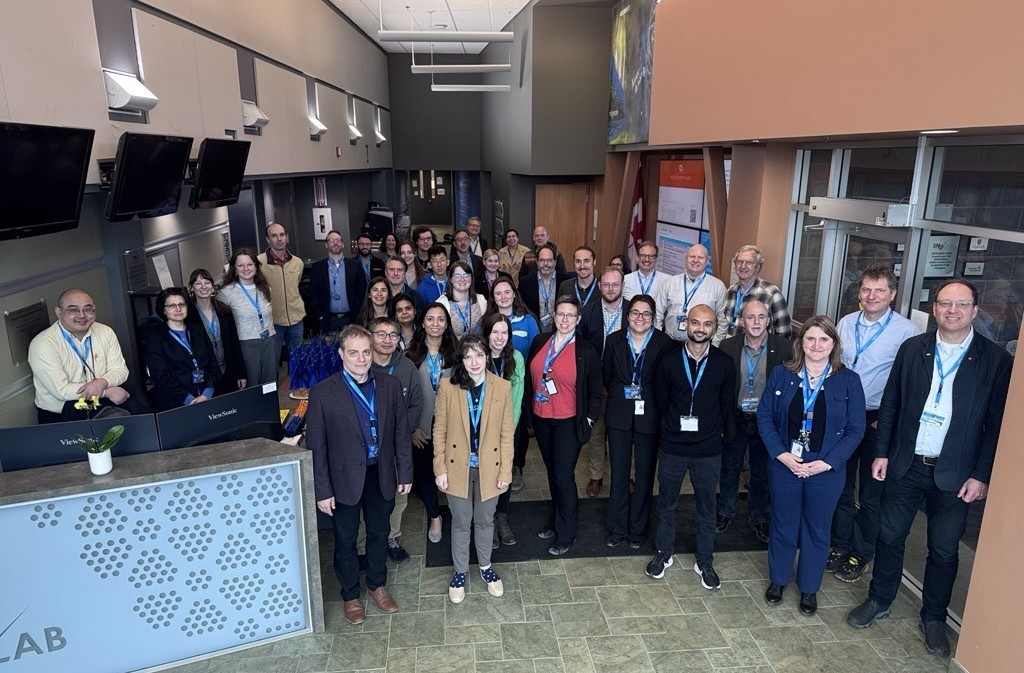}
    \caption{A group photo of the in-person participants at the 2025 SNOLAB Future Projects Workshop. This was taken after the first morning session blocks on day one of the workshop.}
    \label{fig:group_photo}
\end{figure}

The SNOLAB Future Projects Workshop (FPW) is held every two years~\footnote{The 2020 SARS-CoV-2 pandemic interrupted this cycle due to its significant global effects on human health, life, and community.}. The 2025 FPW was held over 2.5 days from April 29 to May 1, 2025.

\begin{figure}[t!]
    \centering
    \includegraphics[width=0.48\linewidth]{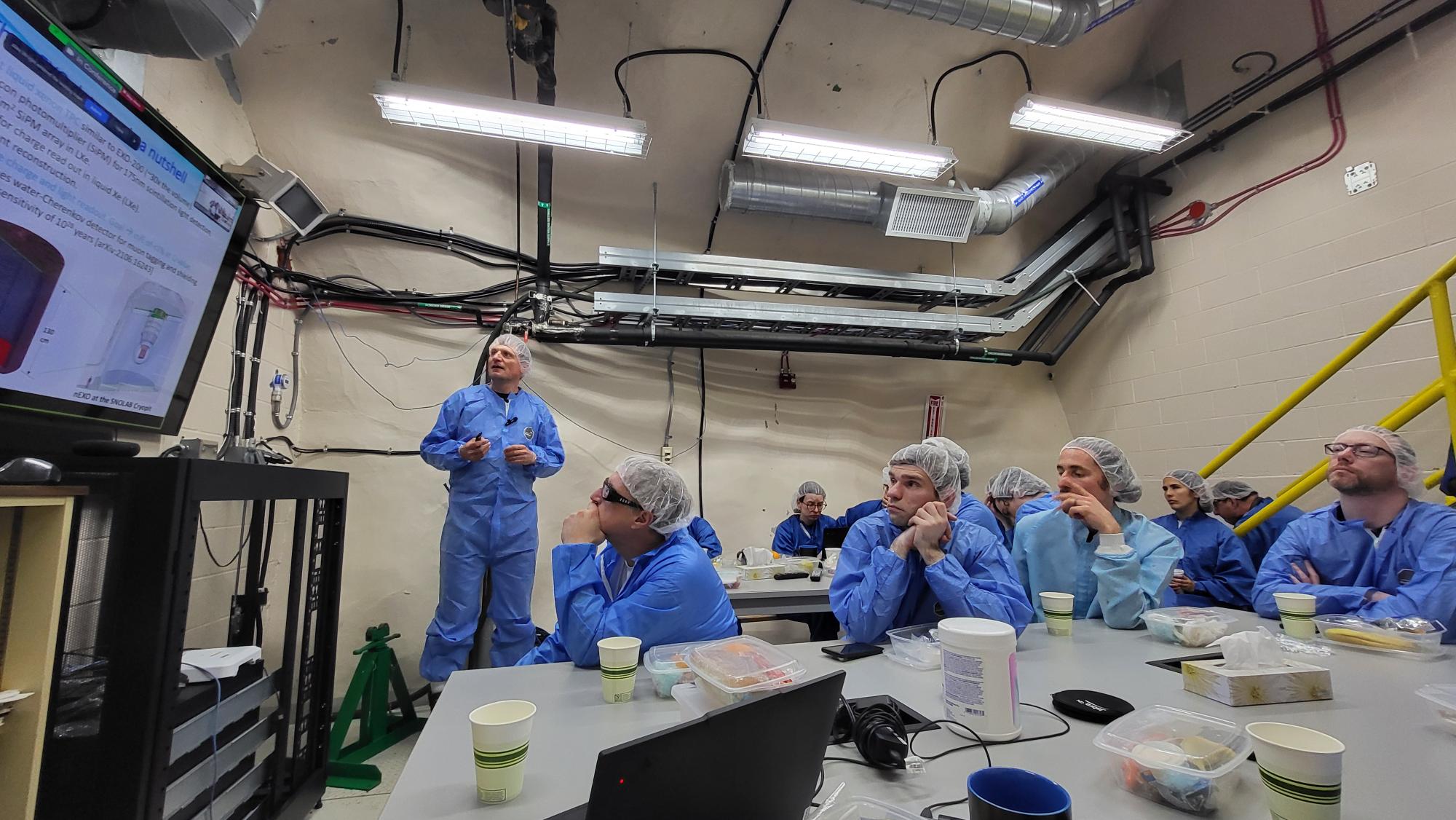}
    \hfill
    \includegraphics[width=0.48\linewidth]{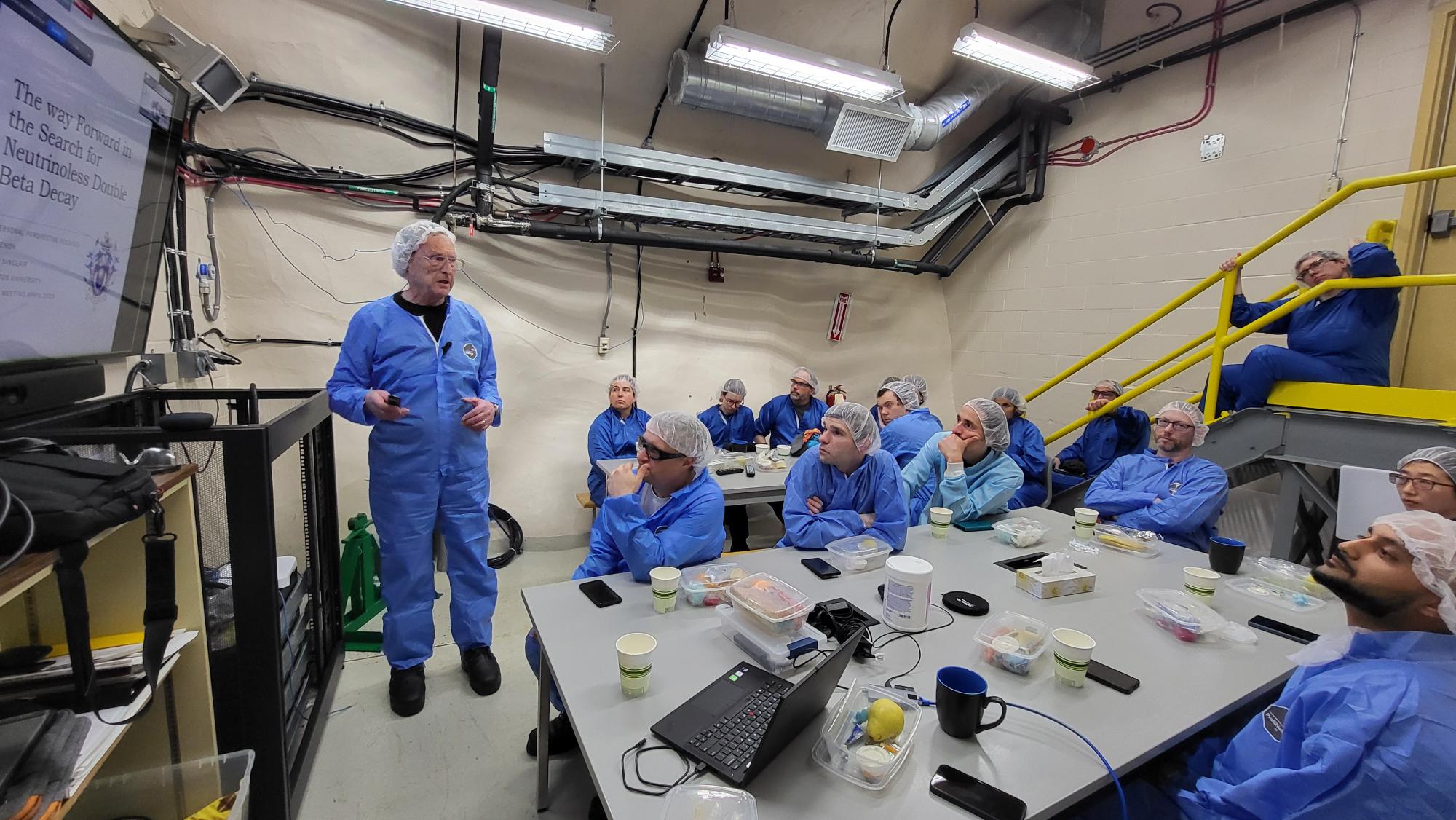}
    \caption{Presentations delivered for the workshop from the underground laboratory.}
    \label{fig:ug_presentations}
\end{figure}

The total engagement in the 2025 FPW consisted of 73 in-person (\Figure{fig:group_photo}) and, at any given time, 20--30 virtual participants. There were 23 presentations by members of the community, two of which were delivered from the underground laboratory itself (\Figure{fig:ug_presentations}). Five of the presentations were given by remote (virtual) participants, three of whom were representatives of SNOLAB's peer deep underground laboratories (DULs): Laboratoire souterrain de Modane (LSM), Laboratori Nazionali del Gran Sasso (LNGS), and the China Jinping Underground Laboratory. Two Canadian funding agencies, the National Science and Engineering Research Council (NSERC) and the Canada Foundation for Innovation (CFI), sent representatives in-person to the workshop. Representatives from the United States Department of Energy connected remotely to listen to parts of the workshop.

The schedule was very full and a few partners or constituencies were unable to give reports. However, this report contains the input that would have been presented in-person, had the schedule permitted. These perspectives are important.

In order to help speakers provide input beneficial to the ongoing 15-year planning exercise requested by CFI, SNOLAB provided some guiding questions in advance:

\begin{itemize}
    \item What do you see SNOLAB’s role for the community to be in the next 15 years? Are there capabilities that you believe SNOLAB will need to host, with accessibility, for the wider community?
    \item How do you see your project fitting into the  field’s landscape on that timescale?
    \item What is the anticipated footprint of your project and what is the most plausible timeline on which you expect to be most active in construction and then operation? What are the space needs in the following three areas: on-site (surface), on-site (underground), and off-site (nearby SNOLAB)?
    \item What are the anticipated on-site personnel levels (separately, in terms of students vs. professionals) during (1) the construction and (2) the steady-state operation phases of the project?
\end{itemize}

\noindent These questions were provided again during the FPW to facilitate group-based brainstorming sessions. Groups were encouraged to appoint a scribe and submit their input to the laboratory directly through an electronic form.

Presentations at the workshop covered a range of topics, from underground biology, to neutrino science (especially neutrinoless double beta decay, denoted $0\nu\beta\beta$ in the rest of this document, and neutrino masses), to approaches to dark matter (DM) direct detection, to novel technologies for gravitational wave detection. After each presentation, and in blocks throughout the schedule, there was time for discussion and questions. 

This report was written by participants in the workshop, as well as other members of the community who could not directly participate but whose perspectives are valuable. Each contributor was asked, at a minimum, to provide 1-2 paragraphs summarizing their material and up to five bullet points containing brief discussions of key points or discussion items. Contributors were free to provide more, of course. These contributions were then edited, the document was approved by the contributors, and a general public-level summary was added. We hope this report will be useful both to the community and to SNOLAB as both embark on multi-year planning activities.

The workshop was organized by a dedicated team of individuals, noted below.

\begin{itemize}
    \item Erica Brunelle (SNOLAB)
    \item David Asner (SNOLAB)
    \item Ken Clark (Queen's University)
    \item Blaire Flynn (SNOLAB)
    \item Pietro Giampa (TRIUMF)
    \item Jeter Hall (SNOLAB and Laurentian University)
    \item Szymon Manecki (SNOLAB and Queen's University)
    \item Nahee Park (Queen's University)
    \item Rachel Richardson (SNOLAB)
    \item Stephen Sekula (SNOLAB and Queen's University)
\end{itemize}

\newpage
\section{Views from SNOLAB and Peer Laboratories}
\subsection{SNOLAB}
\label{sec:topic_snolab}

\authorline{Author: J.~Hall}

\bigskip

\begin{figure}[h]
    \centering
    \includegraphics[width=\linewidth]{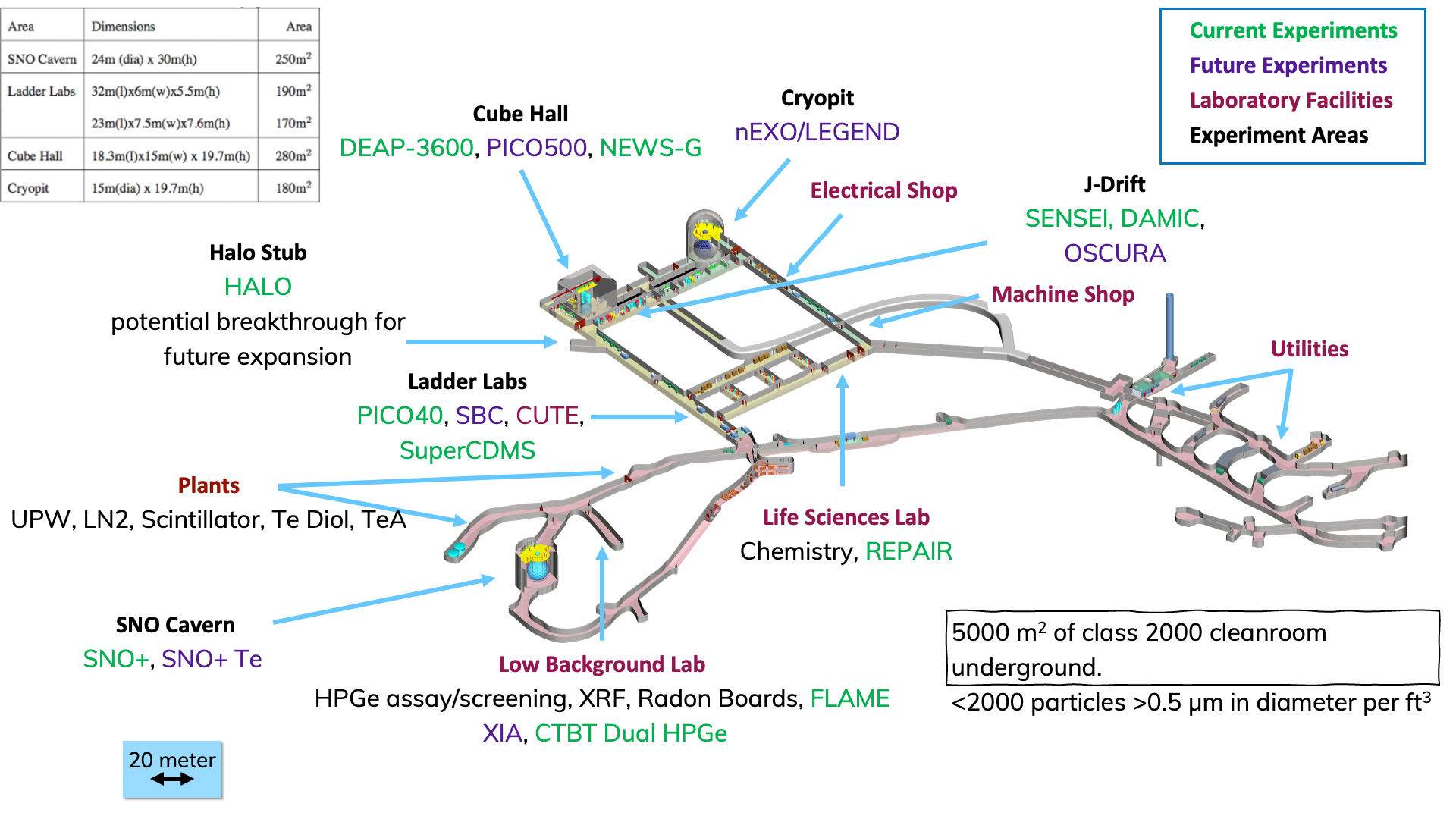}
    \caption{This map shows the layout and location of experiments and facilities in the underground campus at SNOLAB. The potential for future projects is also displayed. This information was valid at the time of the Future Projects Workshop.}
    \label{fig:snolab_map}
\end{figure}

SNOLAB is an underground facility in Lively, Ontario, Canada. The underground campus (Figure \ref{fig:snolab_map}) is located on the 6800~foot (2~km) level in the Creighton Mine operated by Vale International Limited and managed as a joint trust by five universities: the University of Alberta, Carleton University, Laurentian University, Universit\'{e} de Montr\'{e}al, and Queen's University. SNOLAB operations are supported by the Canada Foundation for Innovation and the Province of Ontario~\cite{Smith:2012fq}.

\paragraph{Science Program}

SNOLAB seeks transformational scientific experiments, innovation opportunities and cultural activities that leverage the DUL or the other enabling capabilities at the lab.

The Sudbury Neutrino Observatory was the first project hosted at the current location of SNOLAB. The experimental results from SNO were the basis for Arthur McDonald sharing the 2015 Nobel Prize in physics for the discovery of neutrino oscillations, which show that neutrinos have mass. Studying neutrino properties is still a core part of the neutrino program at SNOLAB. 

In addition to studying neutrino properties, SNOLAB experiments are sensitive to neutrino emissions from industrial and natural sources. These sources allow additional studies of the Sun, the Earth's composition and properties of supernova cores.

SNOLAB's depth allows for the lowest possible radiation environments, by removing the cosmic radiation and through techniques of removing residual radiation from surrounding rock and underground atmospheres. This low background environment allows searches for hypothesized particles that could be masked by the background radiation in surface labs. The most extensively researched new particles are connected with cosmological and dynamical evidence for DM. These DM candidates could interact weakly with baryonic matter, so many DM experiments are hosted at SNOLAB with the goal of detecting various DM candidates.

Other experimental topics that take advantage of the low background environment at SNOLAB include environmental monitoring, quantum computing and sensor development, and low-radiation-dose biological system response.

\paragraph{Enabling Capabilities}

SNOLAB currently employs $\sim140$ personnel including scientists, engineers, technicians, business functions, skilled tradespeople, and cleaners. Lively, Ontario is part of the Greater Sudbury community, which includes a many companies with expertise in deep underground mining and construction. The Creighton Mine is one of the most productive nickel mines operating today.

With a 2~km overburden of norite rock, the total integrated muon flux at SNOLAB of $\sim0.3$~m$^{-2}$~s$^{-1}$ is the lowest in any currently operating DULs~\cite{woodley2024cosmicraymuonslaboratories}. The underground campus includes $5000$~m$^{2}$ of class 2000 cleanroom space to host a variety of experiments.

There is a life sciences laboratory that includes space for growing biological samples, wet chemistry benches, and a fume hood. A small machine shop is operational underground for minor machining and repairs. 

The low background laboratory is utilized for analytical radiochemistry, often measuring long-lived isotopes such as uranium and thorium in construction materials for low background experiments~\cite{Lawson:2023trk}. This laboratory is also used for assay of short-lived isotopes for environmental monitoring. The materials measurement capabilities include several high purity germanium gamma-ray detectors, systems to extract and measure radon, low background alpha-ray detection systems, and an inductively coupled mass spectrometer.

SNOLAB operates the Cryogenic Underground TEst facility (CUTE) as a user facility~\cite{Camus_2024}. This facility includes a dilution refrigerator with both DC and microwave electronics. The inner experimental volume is well shielded with a $\sim10,000$ reduction in ionizing radiation when compared to unshielded surface cryostats.

The ventilation from the mine is capable of removing $\sim$3~MW of heat from the chilled water system that is used to reject heat from the rocks, people, and experiments. The electrical system is currently backed up with 3~MW diesel generators on the surface, which can power the entire underground lab continuously during a power interruption.

The host mining company supplies water to the laboratory, and we host a large water plant to clean up the service water so that it can be used as low background shielding for experiments. The host also provides a compressed air supply, which the laboratory has backed up with an underground compressor system.

\paragraph{International Collaboration}

SNOLAB seeks out both domestic Canadian and international collaborations interested in working at SNOLAB. The SNOLAB user community of $\sim1137$ researchers has a strong component of North American users: 280 from Canada, 428 from the US, and 17 from Mexico. Europe is also well represented with 354 users from fourteen countries. Most of the remaining users are from Asia, although there are a few users from South Africa showing the burgeoning underground science effort in that country.  %

 Projects at SNOLAB are initiated by a formal process that starts with the submission of an expression of interest. The expression of interest is reviewed by SNOLAB management for the impact on the laboratory. In addition to laboratory impact, SNOLAB receives advice from the Experimental Advisory Committee (EAC), made up of an external cadre of scientists with knowledge of the various scientific fields at SNOLAB. Once a project is accepted into the SNOLAB experimental program, it is reviewed biannually by the EAC and by SNOLAB management at important milestones and for any significant hazards. The project management reviews are complete when the experiment enters its operational phase, and the biannual EAC reviews continue until the conclusion of the project. Inquiries about future experiments and projects at SNOLAB can be sent to the Director of Research.

\paragraph{Summary}
\begin{enumerate}
    \item SNOLAB is a DUL in Lively, ON with 2 km overburden of norite rock; the underground campus includes 5000 m$^2$ of class 2000 cleanroom space, and the total integrated muon flux is the lowest in any currently operating DUL. 
    \item SNOLAB’s low background environment hosts searches for hypothetical DM particles, studies of neutrino properties and emissions from industrial and natural sources, environmental monitoring, quantum computing and sensor development, and low-radiation-dose biological system response.
    \item Enabling capabilities at the lab include: life sciences laboratory; small machine shop; analytical radiochemistry, Cryogenic Underground TEst (CUTE) user facility; ventilation and compressed air supply; electrical system with generator backup; and clean water plant.
    \item The SNOLAB user community includes over 1100 researchers from North America, Europe, Asia, and South Africa.
    \item SNOLAB has formal processes for project approval, review, management, and conclusion, to ensure the success of projects throughout their life-cycles.
\end{enumerate}

\subsection{The China Jinping Underground Laboratory}
\label{sec:topic_CJPL}

\authorline{Author: H. Ma}

\bigskip

The China Jinping Underground Laboratory (CJPL) is an underground research facility located in the middle of a traffic tunnel at the Jinping hydropower station in Sichuan Province, southwest China. The traffic tunnel runs through Jinping Mountain, with a rock overburden of more than 1500m along 73\% of its length. The maximum overburden is about 2400~m where the laboratory is located. The first phase of CJPL (CJPL-I) has been in operation since 2010, and the civil engineering of the extension project (CJPL-II) was finished in December 2023 jointly by Tsinghua University and Yalong River Hydropower Development Company.

Two DM direct detection experiments, CDEX and PandaX, as well as GeTHU spectrometers for material screening, are located in the main hall of CJPL-I~\cite{cheng_china_2017}. CDEX uses p-type point contact (PPC) germanium detectors to search for weakly interacting massive particles (WIMPs). The current CDEX-10 phase operates three triple-element PPC germanium detector strings directly immersed in liquid nitrogen in a polyethylene room~\cite{ma_results_2020}. PandaX uses a dual-phase xenon Time Projection Chamber (TPC). PandaX-II operated 580~kg of xenon from July 2016 to July 2019~\cite{cui_dark_2017}. 

\begin{figure}[h]
    \centering
    \includegraphics[width=1.0\linewidth]{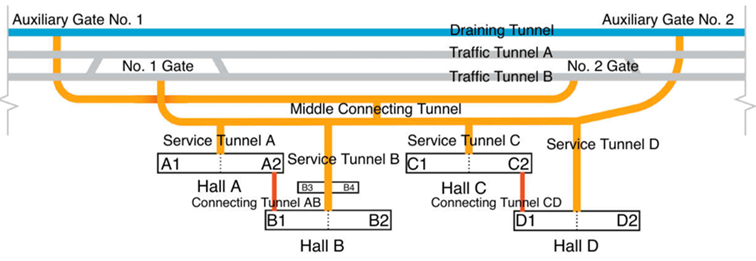}
    \caption{Layout of CJPL-II}
    \label{fig:CJPL-2}
\end{figure}

Two planned-to-be-refilled auxiliary tunnels after the hydropower station were selected for the construction of CJPL-II. The total volume of CJPL-II is about 300~km$^3$ with four main halls of 14~m (width)  14~m (height) $\times$ 130~m (length) shown in \Figure{fig:CJPL-2}. A large pure water vessel is built in hall B2 with a size of 27~m $\times$ 15~m $\times$ 13~m and is used to shield from environmental radioactivity. A large liquid nitrogen tank, located in hall C1 with a diameter of 13~m and height of 13~m, is used to provide shielding and a cryogenic environment for germanium-based experiments. A low background counting facility located in hall C2 consists of an ultra-low background germanium spectrometer with detection limit of 10~$\mu$Bq/kg~\cite{chen_background_2025} and fifteen low background germanium spectrometers with detection limit of 0.1~mBq/kg~\cite{ma_status_2021}. 

Several DM, neutrino and nuclear astrophysics experiments will be located in CJPL-II. For example, CDEX plans to conduct future DM experiment (CDEX-50) and a $0\nu\beta\beta$ experiment (CDEX-300$\nu$) at CJPL-II. These projects aim to set limits on WIMP-nucleon spin-independent coupling cross section down to 10$^{-44}$ cm$^2$ at WIMP masses <10 GeV/c$^2$, and a $0\nu\beta\beta$ decay half-life of $^{76}$Ge to >10$^{27}$~yr, respectively. PandaX is now operating a new TPC with a sensitive target of 3.7~tons of liquid xenon~\cite{zhang_dark_2018}. Several experiments on underground medicine, rock mechanics, and quantum computing have also been proposed to CJPL-II~\cite{ma_status_2021}. They have been approved to have access to CJPL and are now preparing their facilities. 

The Jinping Underground Science Center will be established soon to operate CJPL. The center will be located at the CJPL ground laboratory in Xichang city, about 100~km away from the underground site. In the near future, the center is planned to have about 50 staff, including researchers, engineers, and technicians.

\paragraph{Summary}

\begin{enumerate}
    \item The China Jinping Underground Laboratory (CJPL) is located in Jinping Mountain, Sichuan Province, southwest China, with a rock overburden of about 2400~m, and it is currently the deepest underground laboratory in the world and will become one of the largest with the completion of its extension project (CJPL-II).
    \item In addition to DM, neutrino and nuclear astrophysics experiments, underground medicine, rock mechanics, quantum computing, and other experiments that require a low background environment have been proposed to locate at CJPL.
    \item The Jinping Underground Science Center with about 50 staff will be established soon and dedicated to operating CJPL.
\end{enumerate}%
\subsection{The Laboratori Nazionali del Gran Sasso and the Modane Underground Laboratory}
\label{sec:topic_lngs_lsm}

\authorline{Co-Authors: A.~Ianni, S.~Scorza}

\bigskip

In the following, we discuss briefly the Gran Sasso Laboratory (LNGS) and the Laboratoire Souterrain de Modane (LSM), emphasizing their main research activities and services to support experiments. We also discuss connections between different DULs, in particular within Europe.

\subsubsection{Partnerships}
LNGS and LSM value the strength of collaboration as a driving force behind scientific discovery. We believe that by working together and leveraging the expertise and resources of our global academic and industry partners, as well as all stakeholders, we can achieve greater research excellence, foster innovation, and create meaningful societal impact. We collaborate with other European DULs in the technological developments of infrastructures. We participate in the informal DULIA (Deep Underground Laboratory Integrated Activity) network, formed by the LSM, LNGS, LSC (Laboratorio Subterraneo de Canfranc, Spain), BUL (Boulby Underground Laboratory, United Kingdom), and CallioLab (Finland) laboratories. This network organises workshops on themes related to the research areas of these laboratories. The network works to formalize coordination between its partners at the European level. 

The DULIA network has applied for EU funding within the HORIZON framework to put forward a common program on cross-calibration of instruments for radiopurity assay, new developments in radon reduction and cryogenic technologies, and transnational access. The network has recently been granted funding to exploit a transnational access program which includes laboratories outside Europe, such as SNOLAB, SURF (Sanford Underground Research Facility in the USA), SUPL (Stawell Underground Physics Laboratory in Australia), and a new DUL proposal in South Africa named PAUL (Paarl Africa Underground Laboratory). Agreements between the European laboratories are under development to facilitate staff access and use of facilities.

The LSM is very active in the DUPhy ``Groupe de Recherche'' (GDR)~\footnote{\url{https://gdrduphy.in2p3.fr/}}, as this organisation is at the heart of the coordination of the French efforts in the domain of Deep Underground Physics (DUPhy). The GDR DUPhy was created in January 2021 and is now at its second mandate (renewed in 2025). The GDR DUPhy missions are to ease the development and access of new players at the European underground platforms (e.g. LSM, LNGS, LSC and BUL), to enhance the visibility of the French underground physics community, and to investigate the needs of future experiments in this area. All EU DULs are active in APPEC~\footnote{\url{https://www.appec.org/}}.

In the fall of 2024, the LPSC and LSM jointly submitted a letter of interest to join the ISAPP Network, an alliance of European institutions dedicated to creating a shared curriculum in astroparticle physics at the doctoral level. This initiative aligns closely with LSM's goal of deepening its global partnerships with European and international laboratories, further reinforcing the role of France and IN2P3 in world-leading underground physics research, while actively contributing to the training and development of the next generation of researchers.

The LSM is committed to strengthening collaboration with other DULs through the development of a formal agreement outlining the terms of Research Collaboration and Transnational Access. This initiative reflects the growing synergy and increasing collaborative efforts between the LSM and its international partners.

In line with its mission to educate and inspire future generations, the LSM has established a public museum at its surface facility to enhance outreach and science communication. In a unique blend of art and science, an artistic sculpture now marks the entrance of the surface building—highlighting the lab's dedication to fostering dialogue between physics and the arts. These efforts will continue to be a key part of LSM's outreach strategy.

\subsubsection{Laboratori Nazionali del Gran Sasso}
\label{sec:lngs}

LNGS is the largest DUL in Europe with an excavated volume of 180,000~$\mathrm{m^3}$ and a maximum overburden of 3,800 meters water equivalent (m.w.e.). LNGS is am Italian national laboratory managed by INFN and funded by the Ministry of Research. LNGS currently has 128 staff with 15 researchers, 41 engineers, and 44 technicians. The main research activity at LNGS focuses on astroparticle physics. In addition, research activities on biology, geophysics, and rare processes in fundamental physics are also ongoing. In the next 5-10 years LNGS will be active to support the exploitation of LEGEND-1000 and CUPID in the framework of next-generation $0\nu\beta\beta$ research program. In addition, in the same time window, DarkSide-20k (see Section \ref{sec:darkside}) is entering into operations to search for WIMPs, and XENONnT is being upgraded.

A number of new experimental programs have requested access to LNGS. These are R\&D programs and prototypes for next-generation projects on DM and $0\nu \beta \beta$. They are mainly funded by European grants from the Recovery \& Resilience Facility program \footnote{\url{https://eufundingoverview.be/funding/recovery-and-resilience-facility}}.

LNGS has recently put in operation three new infrastructures. The first is STELLA (Sub Terranean Low Level Assay) equipped with 16 HPGe detectors. The second is NOA (Nuova Officina Assergi), a $400~\mathrm{m^2}$ ISO6 clean room on surface that is designed to be operated radon-free. NOA is currently being used to assembly the photodetectors for DarkSide-20k. Thirs if the Enrico Bellotti Ion Beam Facility for astroparticle nuclear physics.  
LNGS is undertaking an important refurbishment of the main infrastructures: the electrical distribution plant, underground ventilation system, optical link connection between surface and underground, and optimization of the surface laboratory facilities (assembly space, clean rooms, 3D printing workshop, conference room). This activity has been funded by the Italian resilience program. Within the same program new infrastructures are being deployed underground: the cryo-platform, equipped with two cryostats; a new helium liquefier plant with a capacity of 20~L/h, which is replacing the old one with 8~L/h capacity; and a new nitrogen liquefier plant with a cryogenic power of 50~kW at 77~K.

\paragraph{Organization}

LNGS is organized through a Research Division, a Technical Division, and a Coordination Office. The Coordination Office is facilitating the utilization of experiments. Services such as the chemistry laboratory, electronics laboratory, workshop, special techniques (radiopurity assay), and computing are managed by the Research Division. Technical services such as electrical plants, underground facilities and special projects, fluid systems and lifting equipment, and fire and security supervision are managed by the Technical Division. LNGS has also a public affairs and scientific dissemination office and an office for advance training and external funds management and accountability. LNGS has two Functional Units, NOA and an Office for Scientific Strategy, as well as connection with other DULs.  Strong interactions with LSC, LSM, BUL, SNOLAB, and SURF are in place or being discussed.

The lifecycle of experiments at LNGS is managed through a Scientific Advisory Committee (SAC). The SAC analyses the letters of intent (LoI) submitted to the laboratory Director and provides recommendations. The SAC also reviews the CDR and TDR submitted after the LoI approval. In this phase funding agencies are consulted. Technical Services at LNGS support the Director in reviewing the CDR and TDR. Once a project has been approved the building and commissioning phases can start. This activity is managed by the Technical Division. All experiments in operation at LNGS have to define a reference in matter of safety (GLIMOS) and in matter of environmental impact, a technical coordinator, and a site manager to comply with safety procedures at work at the laboratory. Decommissioning plans are also discussed and approved before any experiments start operations.

\paragraph{Experimental Program} LNGS enables a world-class science program. Currently, it has two projects in decommissioning phase, Borexino and DAMA/LIBRA. There are seven projects in the construction/commissioning phase, including DarkSide-20k (see Section \ref{sec:darkside}), COSINUS, and SABRE for DM direct detection. There are ten projects in the running phase, including LEGEND-200, CUORE, CRESST, XENONnT, LVD, LUNA-400, and LUNA-MV. 

The GINGER (Gyroscopes IN GEneral Relativity) project at LNGS aims to probe the Lense-Thirring effect expected from General Relativity at unprecedented sensitivity. The VIP (VIolation of the Pauli exclusion principle) experiment at LNGS aims to probe the Pauli Exclusion Principle (PEP) and investigate the implications of a tiny violation of the PEP. 

Life on the Earth has evolved over billions of years in the presence of natural environmental radioactivity coming from cosmic rays. Understanding whether and how environmental radiation background influences metabolism of living beings is therefore an extremely important aspect in low-dose radiobiology (see also Section~\ref{sec:topic_Deep Underground Biology}). For this reason, useful information can be acquired through the analysis of the differences found in biological systems maintained in parallel on surface and underground (low radiation environment - no cosmic rays). The experiments called PULEX, carried out at the LNGS since the mid-1990s, on cells of different origins (yeast, rodent, human) showed that environmental radiation can act as a stimulus to trigger defense mechanisms against genotoxic damage. The RENOIR experiment, recently funded, aims to understand whether the different biological response observed in underground, with respect to that on the surface, is related to an overall increase of the dose-rate exposure or to the contribution of specific component(s) of the radiation field.

\subsubsection{Laboratoire Souterrain de Modane}
\label{sec:lsm}

LSM is the deepest DUL in Europe. It is supported by CNRS/IN2P3 and the Université Grenoble-Alpes. LSM is an IN2P3 French national platform attached to the Laboratoire de Physique Subatomique et Cosmologie (LPSC) in Grenoble since 2019. It is dedicated to the development of the astroparticle and nuclear physics programs.

\paragraph{Organization} Since 2019, LSM has been a national platform of the French CNRS/IN2P3, hosted and operated by Laboratoire de Physique Subatomique et Cosmologie (LPSC/CNRS and Grenoble-Alpes University). The LSM team (about 10 personnel) is now merged administratively with the LPSC organisation (about 200 personnel) for improved administrative and technical support, along with an enhanced link with the Grenoble University (infrastructure support). The LSM is part of the national strategy on research infrastructure of the French Ministry of Higher Education and Research since 2021.
The activities of the national platform are monitored by a Steering Committee (CODIR). The committee includes the Deputy Scientific Director (DAS) of the ``astroparticles and cosmology'', ``nuclear physics'' and ``interdisciplinary'' fields of IN2P3, and a representative of Grenoble-Alpes University. They meet once a year. As for all IN2P3 national platforms, an External Strategic Council (CSE) is in place to meet once every year to advise the LSM and IN2P3 about its scientific and technology strategies and follow up on its recommendations.

\paragraph{Experimental Program}

LSM is a unique underground laboratory, enabling a world-class science program currently focused on neutrino and DM investigations but expanding to include a broader science base. The program at LSM is attracting internationally renowned scientists and experiments.
While particle astrophysics is the principal focus for the laboratory, there is a long-standing program to host other measurements from other scientific fields that can benefit from the extremely low cosmic-ray background of the LSM and the associated infrastructure for extremely low radioactivity experiments. The LSM has formal agreements for hosting germanium detectors from CNRS groups outside IN2P3 and of CEA, dedicated to environmental sciences. In addition, LSM supports projects studying the impact of underground low radiation environments on biological systems. LSM also has and hosts multiple germanium detectors used to measure background radiation levels in materials.

\begin{figure}[t!]
    \centering
    \includegraphics[width=0.9\linewidth]{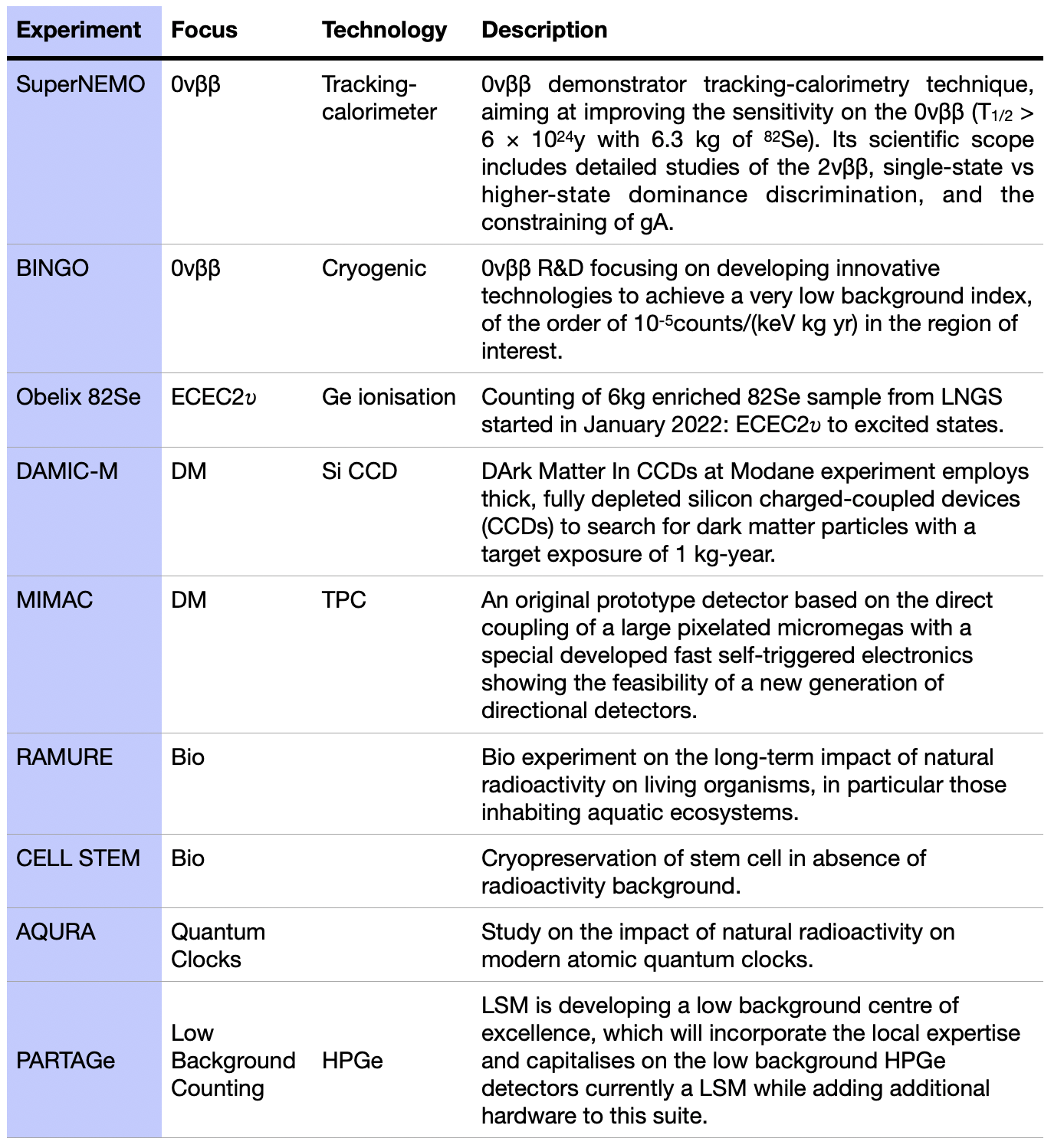}
    \caption{Experiments currently deployed and operational underground at LSM.}
    \label{fig:LSMexp}
\end{figure}

LSM remains firmly committed to delivering a world-class research program (Fig.~\ref{fig:LSMexp}) and advancing the frontiers of underground science. We continue to support key experiments in reaching critical scientific milestones and aim to secure a next-generation flagship project that will define the future scientific direction of the facility.
In parallel, LSM is focused on successfully delivering ongoing projects and completing initial research objectives. Looking forward, we seek to broaden our research portfolio by welcoming new multidisciplinary initiatives that leverage and expand LSM's unique underground capabilities, reinforcing its role as a leading European centre for rare-event physics and related disciplines. A new Expression of Interest (EOI) system \footnote{ \url{https://forms.gle/bh9hR3e13qBhYQG5A}} is now in place for users wishing to propose new experiments or activities underground at the LSM. The EOI system has been active since February 2024 and in the past year the LSM has received nine proposals.

In the area of direct DM searches, the most advanced proposal concerns French group participation in the TESSERACT (Transition Edge Sensors with Sub-eV Resolution And Cryogenic Targets) experiment. This experiment focuses on the detection of low-mass DM (see also Section \ref{sec:cryo_solid_detectors}), with sensitivity ranging from the proton mass down to sub-eV scales, using cryogenic detectors with multiple target materials (Ge, LHe, $\mathrm{Al_2 O_3}$, and GaAs). TESSERACT has received strong support from U.S. research groups and the Department of Energy (DOE), securing funding through the DOE Dark Matter New Initiatives (DMNI) program. It became an IN2P3 Master Project in July 2023 and was presented to the IN2P3 Scientific Council in October 2023. The project was also awarded France 2030 RI2 funding from the French Ministry of Higher Education and Research. The French TESSERACT team plans to install two dedicated cryostats underground at LSM and contribute their expertise in germanium semiconductor bolometers to the experiment's initial detector suite. Still within the field of low-mass DM searches, the DAMIC-M collaboration has expressed interest in continuing R\&D with CCDs at LSM as part of the development of the next-generation OSCURA experiment (see Section~\ref{sec:topic_ccd}). LSM is supporting this proposal, which would strengthen its DM program with two complementary detection technologies.

In the field of $0\nu \beta \beta$, LSM is preparing for the upcoming science results of the SuperNEMO Demonstrator. Looking ahead, there is interest in potential R\&D projects focused on extending the BINGO Demonstrator, reinforcing LSM's strategic role in next-generation rare event physics.

\subsubsection{Summary}

\begin{enumerate}
   \item The \textbf{Laboratori Nazionali del Gran Sasso} (LNGS) is a DUL located in Italy. It is the largest in Europe with 1400 of rock overburden. It is equipped with a number of underground and surface facilities to support research activities. LNGS is developing a strong synergy with other European DULs (LSM, LSC, Boulby, and CallioLab).
    \item Currently, LNGS is undergoing an important renovation process on surface and underground to upgrade the infrastructure and improve  facilities. STELLA and NOA are the first result of this renovation process.
    \item Major experiments are searching for dark matter with XENONnT and CRESST and for neutrinoless double beta decay with CUORE and LEGEND-200. DarkSide-20k and COSINUS are under construction. LEGEND-1000 will be located at LNGS in the Hall C.
    \item The \textbf{Modane Underground Laboratory }(LSM) is the deepest underground research facility in Europe, supported by CNRS/IN2P3 and Université Grenoble Alpes. Since 2019, it has operated as a national IN2P3 platform under the Laboratoire de Physique Subatomique et de Cosmologie (LPSC) in Grenoble, dedicated to the development of astroparticle and nuclear physics programs. 
    \item Shielded by 1,700~m of rock within the Fréjus highway tunnel, LSM provides one of the world’s lowest natural radioactivity environments, ideal for experiments seeking to observe extremely rare physical phenomena.
    \item LSM comprises two main facilities: a 400~m$^{2}$ experimental area hosting international scientific collaborations, and a 50~m$^{2}$ gamma spectroscopy platform, primarily focused on material characterization and validation for ultra-low radioactivity. This facility also supports research in geosciences and biology. The LSM hosts forefront scientific experiments in fundamental physics, including direct dark matter detection and neutrino nature determination—key scientific priorities for the international community in the years ahead.  

\end{enumerate}

\newpage
\section{Small- and Mid-Scale Facilities and Collaborations}

For the purposes of organizing this report, a ``small-to-mid-size" facility or collaboration refers to a project whose physical size can fit in a Ladder-Lab-like space, and/or a collaboration whose size is $\leq 50-100$ people. Size should not be taken as a measure of impact on science, however. The concepts and ideas presented in this part of the workshop could each have a large impact on scientific knowledge.

\subsection{Deep Underground Biology: Past, Present, and Future}
\label{sec:topic_Deep Underground Biology}

\authorline{Co-Authors: M. R. Lapointe, H. Reaume, T. Laframboise, S. Hall, D. Boreham \& C. Thome}

\bigskip

Understanding the role of natural background ionizing radiation (NBR) in normal biological function is a long-standing problem in radiobiology. The first reported experiments in this field took place in the mid-1970s and demonstrated decreased cellular growth rates in low-background environments and accelerated growth in high-background radiation environments~\cite{planel1976demonstration}. With over 30 publications on the topic in the years since, the common observed effects of exposure to sub-NBR environments include reduced growth and/or reproductive fitness, increased genomic stress at baseline, impaired antioxidant defences, and increased sensitivity to subsequent stressors such as acute doses of ionizing radiation or radiomimetic compounds~\cite{planel1976demonstration,lapointe2023protracted,antonelli2008pulex,conter1983demonstration,castillo2021deinococcus,porrazzo2022reduced,zarubin2021first,castillo2015stress,castillo2018transcriptome,fischietti2021low,kawanishi2012growth,carbone2009cosmic,satta2002influence,smith2011exploring,planel1987influence,morciano2018fruit,van2020phenotypic,fratini2015low,satta1995low,luckey1986ionizing}. Furthermore, these effects are often reported to increase in severity with prolonged exposure and tend to return to normal after reintroducing a normal background. Most of these experiments are conducted in DULs, which provide exceptional shielding from the cosmic radiation component of background radiation. However, a significant confounding variable in these works is the variability in dosimetric methods and reporting. These discrepancies render accurate translation of results between laboratories a considerable challenge. To measure the biological impact of this discrepancy, a large-scale, multinational collaborative effort is being organized to run parallel experiments in several DULs to compare the effects of the various low-background environments.

The 2024 DULIA-BIO workshop in York, UK, brought together researchers from around the world to discuss and collaborate on using DULs for biological research. Much of the focus of deep underground biological research revolves around the problem of sub-NBR (or low background) exposure. This meeting led to the formation of a global DUL-Biology collaborative project, which will involve the parallel operation of an identical biological experiment in each of the participating DULs. The DUL-Biology collaboration pilot project will involve a reiteration of the published work by the REPAIR project, implementing a desiccated yeast model~\cite{lapointe2023protracted}. Here, desiccated yeast \textit{Saccharomyces cerevisiae}, prepared by the REPAIR group, will be shipped to each participating DUL. Upon receiving the samples, the receiving group will immediately return a subset of samples to act as transportation controls, then store the remaining samples in their DUL and a control (surface) environment. Every 16 weeks, up to 64 weeks, samples will be collected from each environment and shipped back to the REPAIR group for analysis. Testing will include baseline survival and mutation rates, post-rehydration growth (measured in two ways), post-rehydration metabolic activity, and genotoxic stress analysis. The objective is to build upon the significant results reported by the REPAIR group with novel assays, while reproducing their findings in various low-background environments.

\paragraph{Summary}
\begin{enumerate}
    \item DULs, such as SNOLAB, provide both the location and the scientific, engineering, and logistical support for performing deep underground biology experiments as highlighted at the 2024 DULIA-Bio Conference
    \item The desiccated yeast model experiment provides an opportunity to use a biological model to study the effects of the reduction of natural background ionizing radiation in underground labs with minimal input.  This type of experiment should be some of the first for new laboratories (especially physically smaller labs such as SUPL or the potential PAUL laboratory in South Africa) as they can be started with low input and start generating results within a 6 -- 12 month timeframe. 
    \item More complex models, such as human immune system analogues and organelles, provide a path to continue this type of experimentation over the middle term.
    \item For SNOLAB in particular, the evolution of the biological lab currently in place to a Level III Biohazard safety standard and the possibility of significantly more complex animal models should be considered over the 10 - 15 year range.
\end{enumerate}%

\subsection{HeLIOS -- Detecting Gravitational Waves and Dark Matter with Superfluid Helium}
\label{sec:helios}

\authorline{Co-Authors: M. Hirschel, J. P. Davis}\\

\bigskip

Resonant mass (RM) detectors are using one or more acoustic modes of a low-dissipation solid or liquid medium to resonantly amplify a small periodic force to a detectable displacement signal~\cite{Aguiar2011}. Developed since the 1960s for detecting gravitational waves (GWs)~\cite{Weber1960}, their sensitivity was surpassed in the last two decades by laser interferometers like LIGO. These ultimately succeeded in observing GWs~\cite{Abbott2016}. 

Nevertheless, superfluid helium has recently been suggested as competitive RM detection medium~\cite{Singh2017,Manley2020}, featuring ultra-low dissipation with demonstrated mechanical quality factors exceeding $10^8$ at millikelvin temperatures~\cite{deLorenzo2017}. It also provides the unique possibility of continuously tuning the resonant frequencies through pressurizing the helium, thus effectively broadening the fractional bandwidth of each mode to more than 50\%~\cite{Abraham1970}. A sensitive superfluid RM detector could complement laser interferometers beyond their detection bandwidth (below $\sim 10$~Hz and above $\sim 1$~kHz). On top of that, even a tabletop superfluid RM would act as a promising detector for scalar and vector ultralight dark matter (UDM), surpassing current limits set by laser interferometers after just one hour of integration time~\cite{Vermeulen2021,Abbott2022}. Bosonic UDM is lighter than $\sim 1$~eV and manifests as a wave which is coherent over $\sim 10^6$ periods~\cite{Antypas2022,Derevianko2018}. In the peV-range, it could cause a periodic mechanical signal at low-kHz frequencies (an isotropic strain for scalar~\cite{Arvanitaki2016,Manley2020} or a differential acceleration for vector UDM~\cite{Manley2021}), which would couple to a superfluid acoustic mode~\cite{Hirschel2024}.

In the Davis Lab at the University of Alberta we developed two centimetre-scale proof-of-concept experiments: a prototype superfluid GW detector~\cite{Vadakkumbatt2021} and the Helium ultraLIght dark matter Optomechanical Sensor (HeLIOS)~\cite{Hirschel2024}. The former features a cross-shaped 30~ml helium cavity to maximize coupling to the quadrupolar GW strain, reaching a minimum strain sensitivity of $8 \times 10^{-19}$~Hz$^{-1/2}$ at $\sim$~5.2~kHz and 20~mK temperature. The 145~ml superfluid modes of HeLIOS, once thermal-noise limited at 20~mK, could surpass existing bounds on scalar and vector UDM in the peV-range. Both detectors use a sensitive microwave optomechanical transducer, demonstrate frequency-tunability through pressurization, and promise several orders-of-magnitude improved sensitivity when using reasonably upscaled superfluid cavities.

\paragraph{Summary}
\begin{enumerate}
    \item Besides large flagship experiments, the community should also promote small-scale tabletop experiments. These might exploit novel quantum technologies and explore areas of physics that receive less attention~\cite{Carney2021}.
    \item A superfluid RM detector would not benefit from the overburden of the underground laboratory. However, a collaboration with SNOLAB could enable fruitful knowledge transfer for millikelvin-cryogenic platforms and low-vibration systems. 
    \item The observation of continuous GWs in the kHz-regime (originating from sources such as millisecond pulsars~\cite{Aggarwal2020}) might be of more relevance than peV-range UDM at the moment, in part due to the lack of theoretical motivation for such DM candidates.
\end{enumerate}%
\subsection{Liquid-Noble Bubble Chambers}
\label{sec:lnbc}

\authorline{Author: B.~Broerman}

\bigskip

Bubble chambers are superheated liquid detectors in which particle interactions can create a visible bubble. Because of the low energy required to induce bubble nucleations, these types of detectors are well-suited for DM searches, being especially successful in probing spin-dependent DM-proton cross sections through the PICO suite of experiments using liquid Freon targets~\cite[therein]{Clark:2024nbx}. At nuclear recoil thresholds greater than a few keV, the nucleation mechanism is highly insensitive to bubble formation driven from $\beta/\gamma$ interactions. The remaining $\alpha$-induced backgrounds can be effectively separated from the expected nuclear recoil signature based on acoustic information from the bubble formation. Unfortunately, as the nuclear recoil threshold falls below $\sim$keV, the probability of bubble formation from $\beta/\gamma$'s increases sufficiently to lead to nucleations that, unlike in the case of $\alpha$-induced events, are indistinguishable from nuclear recoils. 

However, noble liquids exhibit a higher $\beta/\gamma$ rejection than in traditionally-used molecular fluids like Freons. The addition of the scintillation mechanism provides an efficient channel for the energy from electron recoils to be expended instead of going into atomic motion, heat, and subsequent bubble formation. Currently only upper limits are set on the bubble nucleation probability from electron recoils in liquid xenon down to a hardware-limited 500~eV threshold while simultaneously remaining sensitive to nuclear recoils~\cite{SBC:2024dub}.

\begin{figure}[!ht]
    \centering
    \includegraphics[width=0.35\textwidth]{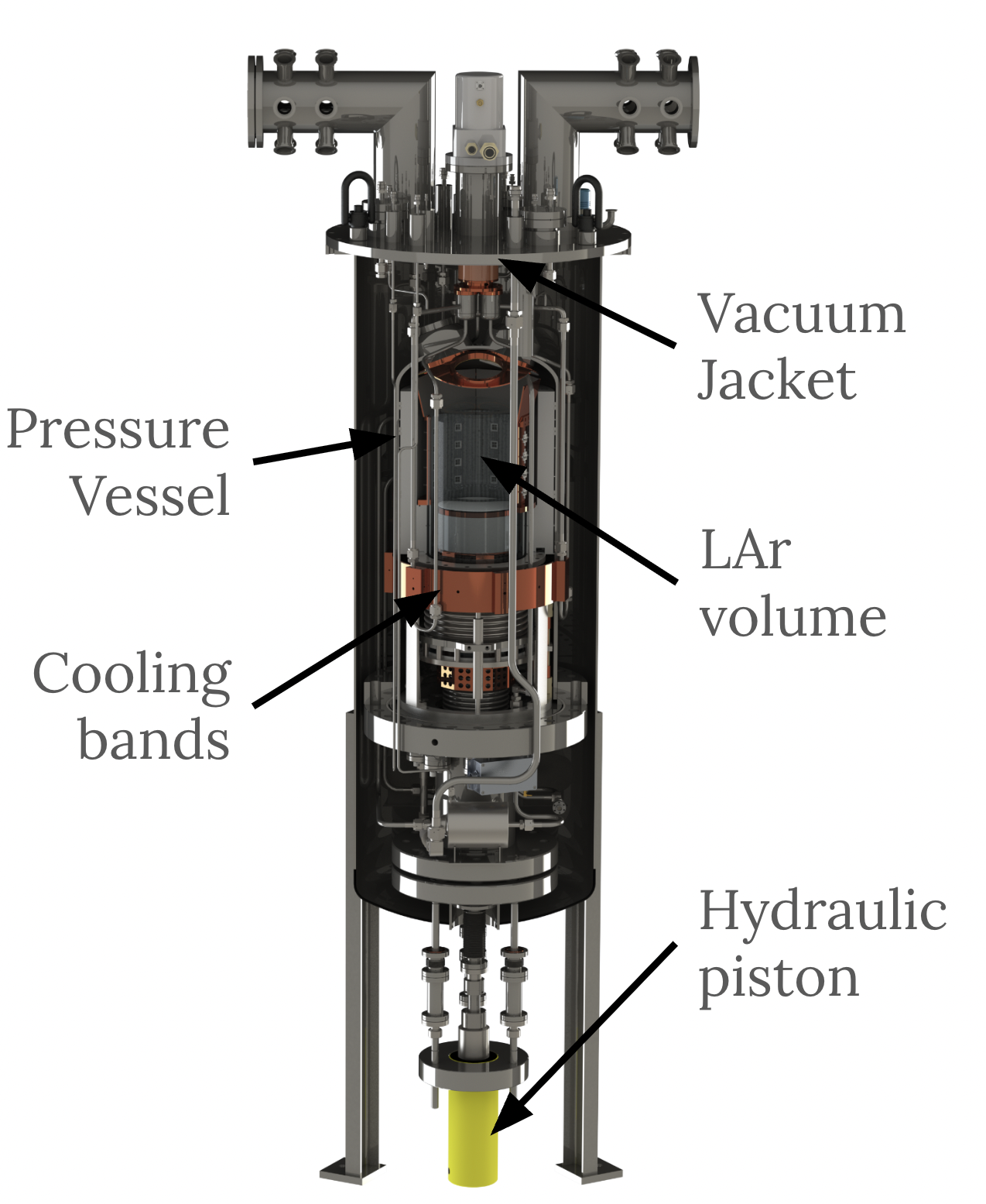}
    \hfill
    \includegraphics[width=0.60\textwidth]{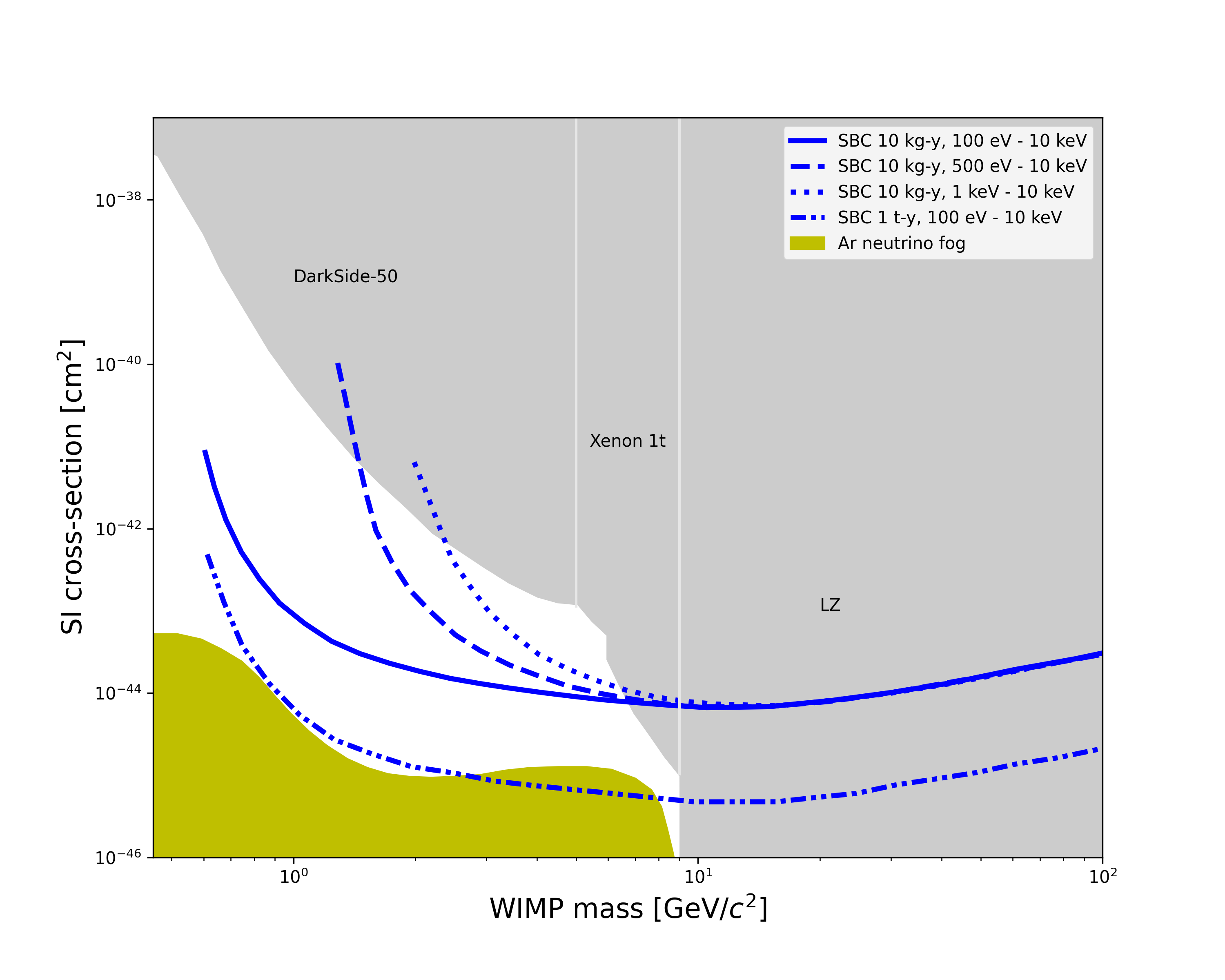}
    \caption{\textit{Left}: CAD rendering of the 10~kg SBC-LAr10 detector to be operated at Fermilab with main component systems annotated. \textit{Right}: GeV-scale DM sensitivity for a 10-kg-yr exposure at several different nuclear recoil thresholds, and the projected 1-tonne-yr exposure assuming only CE$\nu$NS backgrounds. The boundary of the argon neutrino fog is shown in yellow and the currently-excluded regions in gray (from Reference~\cite{Alfonso-Pita:2023frp}).}
    \label{fig:SBC}
\end{figure}

To fully understand the effect of this reduction in electron recoil nucleation probability, the Scintillating Bubble Chamber (SBC) collaboration is building two functionally-identical detectors adopting a buffer-free chamber design~\cite{bressler2019buffer}. The targeted operation (130~K and 30~psi) would allow for a 100~eV nuclear recoil threshold sufficient to probe DM masses in the GeV/$c^2$ mass range~\cite{Alfonso-Pita:2023frp,Alfonso-Pita:2022akn}. In these detectors, 10~kg of liquid argon, doped with xenon to act as a wavelength shifter, is contained in a set of nested, synthetic quartz vessels, and further housed in a pressure vessel, shown in the left of Fig.~\ref{fig:SBC}. Thermomechanical regulation is provided by liquid nitrogen thermosyphons for cooling and an external hydraulic piston for pressure control. The first physics-scale detector, SBC-LAr10, will be commissioned this summer in the MINOS tunnel at Fermilab for engineering and calibration studies and in parallel, assembly at SNOLAB for the low-background DM search detector will begin.

Looking forward to the future, a 1-tonne-yr exposure would allow for a DM search down to the boundary of the argon neutrino fog from $^8$B solar neutrinos, shown in the right of Fig.~\ref{fig:SBC}. A 1~tonne superheated liquid argon volume at 130~K would need dimensions of approximately 1-m-diameter by 1.4-m-high. This volume needs to be pressurized to $\sim350$~psi after a bubble nucleation, and therefore a pressure vessel capable of housing this fluid would require $\lesssim0.01$~mBq/kg of primordial isotopes to generate $<1$ background event/yr from fast neutrons. While these are small activities, they may not be so far off from some upper limits, emphasizing the need for precision assay methods in the future. 

Critically, synthetic quartz vessels, which are traditionally used to contain the superheated liquid, cannot be produced with active volumes greater than 250~L. This provides an opportunity to investigate other detector materials, such as electropolished metals and acrylic, both of which have been demonstrated in small-scale test chambers~\cite{Broerman:2022jtj}. To limit deadtime from repressurizing the detector after a bubble event ($\sim$30~s/event), surface rates from radiogenic and roughness features would need to be kept $<$5~nBq/cm$^2$ to result in $<1$~event/hour. For external shielding, a 9-m-diameter water tank would be sufficient to reduce the fast neutron and Thomson scattering backgrounds from the cavern walls to $<1$~event/year each.

Liquid-noble bubble chambers are a promising technology sensitive only to nuclear recoils in the GeV-scale DM region of interest. Additionally, there is flexibility in the type of active fluid that can be used (Xe, N$_2$, CF$_4$, etc.), allowing for both an expanded physics reach (e.g.~to spin-dependent searches) and the quick cross checking of potential positive signals if observed. To fully explore this parameter space, R\&D will be required to tackle the technical challenges in scaling up to a single 1-tonne volume, or 500~kg volume operating for 2~years to achieve tonne-yr exposures. 

\paragraph{Summary}
\label{sec:sbc_summary}

\begin{enumerate}
    \item We should not discount discovery in the search for DM as we reach towards the neutrino fog. There is still parameter space in the traditional WIMP regions to close out with viable, even predicted masses. 
    \item As we continue, and possibly expand, operations underground, we will need more resources (staff, research scientists, consumable materials, hardware, etc.) and logistical on-site support (transportation underground, preparation spaces on surface, assaying services). The shielding requirements for proposed experiments may need water tanks that will not fit in the current ladder labs. 

\end{enumerate}%

\subsection{Cryogenic Solid-State Detectors: SuperCDMS and Beyond}
\label{sec:cryo_solid_detectors}

\authorline{Co-Authors: M. Diamond, Z. Hong}

\bigskip

In the following we summarize the SuperCDMS SNOLAB experiment, and R\&D for possible future cryogenic solid-state detector experiments using similar technology.

\paragraph{The SuperCDMS SNOLAB experiment,} currently under construction, will primarily target nuclear recoils of 1--10~GeV WIMP-like DM particles. It will also search for electron recoils of WIMP-like particles at sub-GeV masses, and absorption by electrons of dark photons, axion-like particles, and lightly-ionizing particles at the eV--keV mass scales. It utilizes cryogenic calorimeters cooled in a dilution fridge. These consist of an array of detectors, each with a kilogram-size absorber of ultra-pure Si or Ge crystal~\cite{SCDMS_SNOLAB}. The key to the detectors' energy resolution lies in the Transition Edge Sensors (TESs) on the top and bottom crystal faces, which are weakly linked to the thermal bath and measure the athermal phonon signals. There are two types of detectors in the array, namely interleaved Z-sensitive Ionization and Phonon (iZIP) designs for discrimination of nuclear vs. electron recoils, and High Voltage (HV) designs that reaches lower energy thresholds while sacrificing this discrimination ability. The iZIPs feature charge-sensing electrodes amplified by HEMTs (High Electron Mobility Transistors) interleaved with the TESs, to provide dual ionization and phonon readout; the ratio of the energies collected in these two readout channels on an event-by-event basis permits the recoil discrimination. The HVs, with only phonon readout, rely on the Neganov-Trofimov-Luke (NTL) effect to generate a large number of phonons by drifting charge carriers through a potential, generated by up to 100~V applied across the crystal.

Vibration and seismic isolation, multiple layers of shielding, radiopurity of materials, radon reduction, and cavern overburden all play a role in background reduction. 
The residual nuclear-recoil background will be dominated by the irreducible contributions from solar neutrinos.
The initial payload, $\sim 30$~kg of crystal, will consist of 4 towers of 6 detectors each. Installation and integration activities are currently ongoing, with the first full science run expected to begin in early 2026. One of the HV towers was operated in the CUTE (Cryogenic Underground TEst) facility Oct 2023 -- March 2024, including $\sim 2$~months of calibration runs, $\sim 2$~months of low-background runs, and several weeks dedicated to detector characterization such as noise assessments. While the science exposure was small, the critical mission was to exercise and debug the detectors before SuperCDMS cryostat is in place, in a similar environment and with the same electronics. 

SuperCDMS SNOLAB is expected to operate through at least 2028, and long-range planning exercises by the collaboration~\cite{SCDMS_LRP} have already scoped out the potential sensitivity of another subsequent generation of the experiment. Pushing the sensitivity to even lower DM masses will require further improvement in phonon resolution, and further reduction or subtraction of low-energy backgrounds. Options for pushing the sensitivity down in cross-section, into the ``neutrino fog'', include improving recoil discrimination via iZIP detectors with better ionization resolution, or via ``piZIP'' detectors that would measure recoil phonons separately from NTL phonons.

\paragraph{HVeV} detectors, prototype gram-scale Si devices with eV-level resolution, are at the core of an ongoing R\&D program to probe the limits of TESs and understand particular excesses that show up in the low-background run spectra~\cite{SCDMS_HVeV}. Peaks at energies lower than the single electron-hole pair (1~eh) ionization energy may come from charge injections or trapping; such events may be more likely to occur at the unpolished curved sidewalls of the crystals. An unexpectedly large peak at the 1~eh energy may come from charge leakage at the electrodes, or light leakage into the detector enclosures. An exponentially-falling ``low-energy excess'', consistent with excesses seen in various other experiments, appears to be heat-only (i.e. non-ionizing) events. The latest HVeV run was conducted at CUTE \cite{SCDMS_HVeV_CUTE}; as well as the low-background facility, the setup featured better light tightness and quieter electronics than previous runs. The HVeVs themselves featured more sensitive TESs, and some incorporated an SiO$_2$ insulating layer to mitigate charge leakage backgrounds. The HVeVs were operated at multiple voltages to inform models of the non-ionizing excess. HVeVs are now being fabricated with different crystal materials, including Ge, SiC, and diamond; these may also be tested in SNOLAB in the future.

\begin{figure}[h]
    \centering
    \includegraphics[width=0.9\linewidth]{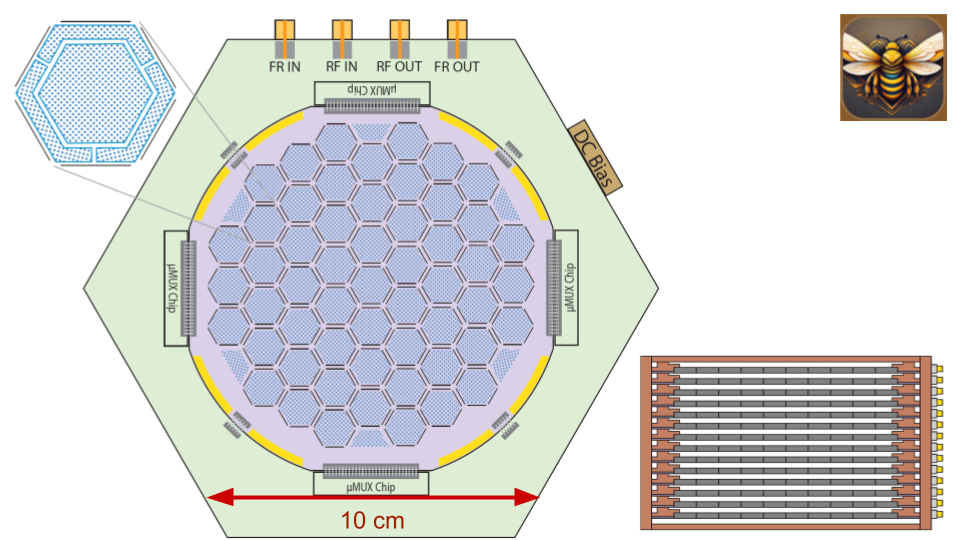}
    \caption{Conceptual design of HONEYCOMB: HVeV-like detectors (top-left corner) combined into wafers (center) stacked on top of each other (bottom-right corner).}
    \label{fig:honeycomb}
\end{figure}

\paragraph{HONEYCOMB} is a plan to scale up the HVeV concept by combining many detectors, weighing a few grams each, into honeycomb-shaped layers with $\sim 10$~cm faces (see Fig.~\ref{fig:honeycomb}). Deep Reactive Ion Etching would be employed to create trenches in each wafer of crystal material to form the honeycomb. This would leave a small ``bridge'' for thermal conduction between hexagonal areas in the honeycomb; the smooth dicing would reduce sidewall charge injection. Stress-free mounting would be employed, with the goal of mitigating the low energy excess by reducing events caused by crystal strain. Multiplexing readout with $\mu$MUX, with multi-stage cryogenic amplifiers, would be required. Multiple wafers would be stacked on top of each other, for a total payload of $\mathcal{O}(10)~\mathrm{kg}$, with the wafers serving as active vetoes for each other. The energy threshold goal is at the eV level. Recent RF readout upgrades at CUTE make this facility an ideal host.

\paragraph{SPLENDOR} is a plan to penetrate the neutrino fog using directional detectors, based on narrow-bandgap materials that have anisotropic crystal structures. The collaboration has grown Eu$_5$In$_2$Sb$_6$ ($\sim 40$~meV indirect bandgap) and Eu$_5$Zn$_2$P$_2$ ($\sim 10-100$~meV indirect bandgap). Low-threshold ionization readout would take full advantage of the low bandgap of these materials. An R\&D setup using two-stage HEMT readout has already achieved 7.2~electron resolution~\cite{SPLENDOR}.

\paragraph{Summary}
\label{sec:cryoSD_summary}

\begin{enumerate}
\item Commissioning of SuperCDMS SNOLAB is currently underway, with science running expected 2026 through 2028.
\item The low energy threshold of SuperCDMS detectors enables low-mass nuclear recoil DM searches; iZIP technology provides background rejection while HV technology pushes down the sensitivity threshold.
\item SNOLAB provides critical resources for all aspects, including detector testing at CUTE.
\item Si HVeV (gram-scale prototype) detectors achieve sub-1~eV resolution. More HVeV varieties are coming soon.
\item Potential future experiments include HONEYCOMB (extreme phonon sensor based on scaled-up HVeV design) and SPLENDOR (charge-readout directional detector using narrow-bandgap anisotropic crystals).
\item For future experiments limited by tritium backgrounds (due to cosmogenic activation), underground detector fabrication and copper electroforming could be greatly beneficial. 
\item There is the possibility of sharing infrastructure with other future cryogenic solid-state detector projects, e.g. Globolo (see Section \ref{sec:globolo}), and synergy with TESSERACT (see Section \ref{sec:lsm}).
\end{enumerate}%
\subsection{Skipper-CCD and Related Efforts}
\label{sec:topic_ccd}

\authorline{Co-Authors: J.~Estrada, A.~Chavarria, J.~Tiffenberg}

\bigskip

Over the past decade, Charge-Coupled Devices (CCDs) and their advanced variant, skipper-CCDs, have emerged as transformative technologies in the field of direct DM detection. SNOLAB has played a critical enabling role on this since 2012. The remarkable sensitivity of CCDs to minute ionization signals enables the detection of single or few-electron events, a capability well-suited for probing DM candidates with sub-GeV masses. The DAMIC~\cite{damic2020PhRvL.125x1803A, damicskipper_2024PhRvD.109f2007A} (Dark Matter in CCDs) experiment, operating at SNOLAB, first demonstrated the potential of high-resistivity CCDs to operate in low-background environments, achieving exceptional thresholds and radiopurity. Building upon this foundation, SENSEI~\cite{sensei2025PhRvL.134a1804A, senseiPhysRevLett.134.011804} (Sub-Electron-Noise Skipper CCD Experimental Instrument) introduced skipper-CCDs, which allow for repeated non-destructive readout of each pixel, reducing noise to below a single electron and unlocking new parameter space in the search for DM-electron interactions. The DAMIC-M experiment~\cite{damicM}, in construction now at LSM in France (Section~\ref{sec:lsm}) with participation from the SNOLAB team, integrates these advancements into a large-mass, low-threshold instrument optimized for rare event searches. Together, these experiments have redefined the frontier of low-mass DM sensitivity, demonstrating the power of precision silicon sensors in uncovering the fundamental nature of DM.

The long-term goal of the direct DM skipper-CCD program is Oscura~\cite{oscuraexperiment, OscuraSensors2023JInst}. This project will realize the full potential of skipper-CCD technology in the search for DM by scaling up to a 10~kg target mass. This project has finished its design stage as part of the Department of Energy's Dark Matter New Initiatives portfolio. Oscura represents a community-wide effort to build a next-generation skipper-CCD detector at SNOLAB. This ambitious project unifies the expertise and infrastructure developed through DAMIC, SENSEI, and DAMIC-M, bringing together the entire skipper-CCD DM community to address the remaining scientific and technological challenges in sensor fabrication, ultra-low background operation, and high-throughput readout. 

\paragraph{The currently running SENSEI experiment} has produced world-leading results in low mass DM searches. We expect this experiment to continue operating with a larger detector mass approaching 100~g in the next couple of years. After this next run is complete, the SENSEI infrastructure will be leveraged for testing new sensor technologies at low background. The technologies currently under consideration are CCD sensors with low noise and faster readout (MAS-CCD, SiSeRO-CCD and skipper-CMOS), and with double-sided readout.

\paragraph{The original DAMIC detector at SNOLAB} has become a test stand for background studies. DAMIC is proposing an upgrade to perform nuclear-recoil versus electron-recoil discrimination by identifying the defects left behind in the silicon crystal by nuclear recoils \cite{DAMICdiscr_McGuire:2023cbg}. Once the run with this upgraded capability is complete, the infrastructure could be used for further development of background-suppression techniques.

\paragraph{Oscura} is technically ready to begin construction, but a clear funding path for the full-scale experiment has not yet been secured. In the interim, the Oscura team is building a 1-kg skipper-CCD payload at Fermilab, which serves as a full integration test of the technologies required for the larger experiment. This prototype will be operated in a an existing cryogenic vessel at the Fermilab surface facility, providing critical validation of the detector performance and readout infrastructure.

The next phase of the plan involves constructing a low-background cryostat at SNOLAB, designed to operate the 1-kg payload in a deep underground environment~\cite{Oscura_OIT2024}. Importantly, this vessel will also be engineered to support future upgrades to a 10-kg payload, enabling a seamless transition to the full Oscura experiment. New collaborators from Canada --- specifically Yoni Kahn at the University of Toronto and Ana Botti at the University of Montreal --- have expressed strong interest in contributing to this effort. 

\paragraph{Summary}
\begin{itemize}
\item By delivering unprecedented sensitivity to sub-GeV DM, Oscura will take skipper-CCD-based direct detection to its full potential.
\item A successful demonstration of the 1-kg payload will position the Oscura collaboration to make a compelling case for international funding of the full-scale experiment, leveraging support from the U.S., Canada, and additional partners.
\item Existing SENSEI and DAMIC infrastructure can be leveraged for various aspects of the Oscura effort.
\item There would be obvious collaboration opportunities between Oscura and the low-mass DM detection ideas that will be presented in Section~\ref{sec:SmallScaleDM}.
\end{itemize}%
\subsection{Novel Small-Scale Experiments for Light Dark Matter}
\label{sec:SmallScaleDM}

\authorline{Co-Authors: Y. Kahn, B. Roach, N. Hoch, A. Williams, E. Marrufo Villalpando, D. Baxter, A. Drlica-Wagner, P. Abbamonte, D. Balut, J. Oh, L. Thompson, D. Johnson, D. Freedman, L. Winslow, C. Blanco, A. Radick, B. Lillard, C. Gao, J. Sch\"{u}tte-Engel, W. Halperin, M. Nguyen, J.W. Scott, J. Foster}

\bigskip
Here we describe two small-scale DM experiments which are highly synergistic with SNOLAB's existing programs (e.g. CCDs as described in Section \ref{sec:topic_ccd}) : a search for sub-GeV DM-electron scattering with organic scintillator crystals coupled to CCDs, and a search for neV-mass axions coupled to nuclear spins with superfluid ${}^3{\rm He}$ and quantum sensors.

\paragraph{Organic scintillator crystals} such as trans-stilbene (C$_{14}$H$_{12}$) are excellent targets for DM-electron scattering~\cite{Blanco:2021hlm}, for several reasons: the electronic excitation energy is a few eV, the de-excitation yields a slightly wavelength-shifted photon to which the crystal is transparent, the crystals can be grown to macroscopic sizes, and the molecular structure is naturally anisotropic which results in daily modulation of the event rate. Existing collaborative efforts at multiple institutions have shown that trans-stilbene can be grown in gram-scale single crystals (MIT); the scintillation light can be read out with a back-thinned skipper CCD (Fermilab); and the anisotropic response of the crystal to electron scattering is potentially measurable from the scintillation photon flux (MIT and University of Illinois). Ongoing theory work at the University of Oregon and Penn State University has determined optimal crystal orientations in order to maximize the daily modulation signal. If a prototype can be demonstrated to have sufficiently low-background rates in a surface run, a 100-gram-scale detector (possibly consisting of several related target compounds, each with its own anisotropic response and scintillation spectrum) could achieve leading sensitivity to DM-electron scattering in the MeV-GeV mass range, with the daily modulation providing a handle to evade the ubiquitous low-energy excess.

\paragraph{Ultralight axions} are an appealing theoretical candidate for DM, but the extremely weak coupling to matter and the fact that the axion mass is unknown typically motivate experiments which can boost the signal-to-noise ratio through resonance. One example is an NMR-type experiment (for example, CASPEr~\cite{JacksonKimball:2017elr}) where the tip angle of nuclear spins resonantly increases when the Larmor frequency is tuned to the axion mass. A recent proposal by Kahn and collaborators~\cite{Gao:2022nuq,Foster:2023bxl} showed that superfluid ${}^3{\rm He}$ at mK temperatures in the B-phase can naturally scan through many axion masses at once: when prepared in the homogeneous precession domain (HPD) with an RF pulse, relaxation on a timescale $T_1 \sim 1-1000~{\rm s}$ will result in a slowly-varying Larmor frequency in the MHz range~\cite{bunkov2012spin}, corresponding to neV axion masses. The axion interaction with the helium spins will result in a tiny ``blip'' in the precession frequency drift, with a fractional size of $10^{-13}$. A measurement this sensitive can best be achieved with quantum sensors, specifically superconducting qubits locked to a frequency standard in an optimal quantum control scheme for frequency measurement~\cite{naghiloo2017achieving}. While several experiments have been proposed to search for this coupling of the axion to nuclear spins, there have been no laboratory improvements on astrophysical limits in this mass range for decades~\cite{Berlin:2024pzi}. This proposed experiment combines an appealing science target with a concrete proposal to use quantum sensing techniques for fundamental physics.

\paragraph{Summary}
\label{sec:smallscaleDM_summary}

\begin{enumerate}
    \item Organic scintillator crystals such as trans-stilbene coupled to skipper CCDs are a viable path forward to scale the target mass for sub-GeV DM experiments while offering the additional handle of daily modulation. Such an experiment builds on the success of SENSEI and is highly complementary to the planned Oscura program (Section~\ref{sec:topic_ccd}).
    \item Superfluid ${}^3 {\rm He}$ in the homogeneous precession domain of the B-phase can be used to detect axion DM; readout using superconducting qubits and optimal quantum control can help bridge SNOLAB's investment in quantum science with its main mission of fundamental physics research.
    \item Both experiments could benefit from the technical expertise of SNOLAB staff: organic chemistry for the sub-GeV experiment and cryogenics for the axion experiment.
\end{enumerate}%
\subsection{Bolometric Detectors: Precision Beta Decay and Future Neutrinoless Double Beta Decay Searches}
\label{sec:globolo}

\authorline{Co-Authors: K.~J.~Vetter, L.~A.~Winslow}

\bigskip

Bolometric detectors, especially those based on cryogenic calorimetry, are uniquely suited for a distributed network approach to reach meV-scale $m_{\beta\beta}$ sensitivity, which may be required for $0\nu\beta\beta$ discovery in the normal ordering scenario. Current-generation technologies such as those used in CUORE~\cite{cuore2022search} have demonstrated stable operation of a tonne-scale segmented detector over a multi-year timescale. CUPID~\cite{cupid2023} will build on this heritage with enriched scintillating bolometers, combining heat and light readout to provide powerful $\alpha$ background rejection. This technique has already been demonstrated by the CUPID-Mo~\cite{augier2022final} and AMoRE-I~\cite{agrawal2025improved} experiments. The inherent modularity of bolometric detectors allows the community to gradually scale up exposure and sensitivity in parallel with tightening constraints from cosmology, direct mass measurements, and oscillation data.

The future of bolometric detectors may lie in distributed cryostats, which could collectively achieve the 10--100 tonne$\cdot$year exposures needed for meV-scale sensitivity. This approach requires advances in quantum sensing technologies and sophisticated sensor readout systems. Concepts such as CUPID-1T envision tonne-scale, multi-cryostat calorimetric detectors located at multiple underground locations that together could reach half-life sensitivities approaching $10^{28}$ yr and probe effective Majorana neutrino masses in the 4--7 meV range, providing discovery potential within the normal ordering parameter space~\cite{collaboration2022toward}. The advantages of a distributed network include:

\begin{itemize}
    \item Utilization of commercially available cryogenics that support automated, scalable operation
    \item Increased robustness, as portions of the network can continue running independently if others experience downtime
    \item Reduced technological risk through diversification -- different sites could implement complementary detector technologies, much like collider experiments
    \item Rapid cross-confirmation of potential discoveries through observations at multiple locations
\end{itemize}
\noindent 
Disadvantages include increased costs and more personnel needs.

A variety of advanced detector technologies are under development for the next generation of $0\nu\beta\beta$  searches, including magnetic microcalorimeters (MMCs), kinetic inductance detectors (KIDs), and neutron transmutation doped (NTD) thermistors. TES-based readout has been a particular focus of the U.S. CUPID group, with surface demonstrations of light detectors~\cite{singh2023large} and cryogenic Li$_2$MoO$_4$ detectors~\cite{bratrud2025first} indicating potential for future $0\nu\beta\beta$ searches. For large-scale detectors, thousands of channels will be read out simultaneously, thus it will be necessary to use multiplexed readout hardware to reduce noise and hardware costs. Multiplexing is an active area of development for $0\nu\beta\beta$  searches and other applications. These developments complement other efforts using TES for rare event searches and precision measurements, such as the TESSERACT~\cite{chang2025first} and HOLMES~\cite{borghesi2023updated} experiments, creating a diverse ecosystem of cryogenic detector technologies that can be optimized for different scientific goals.

In the near term, precision $\beta$ decay measurements provide a crucial pathway to validate detector performance and refine nuclear theory inputs. TES-instrumented detectors deployed at facilities like SNOLAB's CUTE can perform high-resolution spectral measurements of $^{115}$In $\beta$ decay, helping discriminate between theoretical nuclear models and improving nuclear matrix element calculations for $0\nu\beta\beta$ ~\cite{pagnanini2024simultaneous, leder2022determining}. These measurements represent an essential R\&D step toward the ambitious goal of a global bolometric network capable of probing the most challenging parameter space in neutrino physics.

\paragraph{Summary}
\begin{enumerate}
    \item A distributed network of tonne-scale scintillating bolometric experiments provides a realistic and scalable path toward meV-scale sensitivity to $m_{\beta\beta}$, with the flexibility to expand as global constraints on neutrino parameters evolve. The distributed network is a practical alternative to monolithic detector designs, which may face higher upfront infrastructure costs and engineering challenges at larger exposures.
    \item Bolometric technologies are mature and versatile, offering sensor options such as NTD-Ge thermistors, MMCs, KIDs, and TES, which enable optimization for timing resolution and background rejection.
    \item Sensor R\&D must continue to achieve better detector resolutions and lower thresholds, and multiplexing hardware must be tested to ensure the feasible readout of thousands of bolometer channels.
    \item Near-term precision beta decay measurements with TES-instrumented crystals can validate detector performance and provide critical input for nuclear theory.
    \item SNOLAB, through the CUTE facility, is well-positioned to play a key role in early R\&D and could serve as one of the sites in a future global bolometric network.
\end{enumerate}%

\newpage
\section{Large-Scale Facilities and Collaborations}

For the purposes of this report, a ``large-scale" facility or collaboration refers to a project that requires a dedicated cavern-scale space to be hosted (in its final envisioned phase), and/or a collaboration of $\geq 100$ people. The size and complexity of these projects generally stems from the requirement for large exposure, necessitating large mass or volume, to reach the target sensitivity.

\subsection{High Energy Neutrinos}
\label{sec:HEneutrinos}

\authorline{Author: C. B. Krauss}

\bigskip

The IceCube observatory's~\cite{IceCube-Gen2:2020qha} evidence of extragalactic sources of neutrinos and the recent measurement of neutrinos from the Milky Way mean that high-energy neutrino astronomy is quickly becoming a key new tool in the arsenal of scientific observatories. This applies to both particle physics, giving the field access to  particles at the highest energies, and to multi-messenger studies that will yield new insights into nature. This reality was recently reinforced by the observation of an extremely high energy particle at the ORCA site of KM3NeT \cite{Aiello2025}.

\paragraph{Summary}

\begin{enumerate}
    \item High-energy neutrino observatories are instrumental for uncovering more of the origins of the highest-energy processes in the universe. Due to the opacity of the Earth to neutrinos at energies higher than 60~TeV, several telescopes are needed to achieve full neutrino sky coverage. The Pacific Ocean offers an excellent location with a good depth of 2,600~m, good optical transparency, a fully instrumented site and excellent complementarity to the European and existing Asian sites in the Mediterranean and Lake Baikal, respectively. 
    \item P-ONE has been proposed for the Cascadia Basin site of Ocean Networks Canada's Neptune underwater network. A demonstrator that will be able to conclusively prove the site's scientific viability has been funded in Canada and with international partners~\cite{Agostini2020, Bailly2021}.
    \item SNOLAB has valuable experience when it comes to project management, project engineering, optical sensor calibration, and has a group of scientists already supporting neutrino projects. Contributions to P-ONE would be similar in nature and impact as the TRIUMF laboratory's current contributions to P-ONE, where the experiment benefits from site infrastructure in Vancouver and the data acquisition system of P-ONE is largely a TRIUMF deliverable. 
\end{enumerate}%
\subsection{SNO+ Tellurium Double Beta Decay}
\label{sec:topic_SNO+}

\authorline{Co-Authors: M.~Chen, for the SNO+ Collaboration}

\bigskip

\begin{figure}[h]
    \centering
    \includegraphics[width=1.0\linewidth]{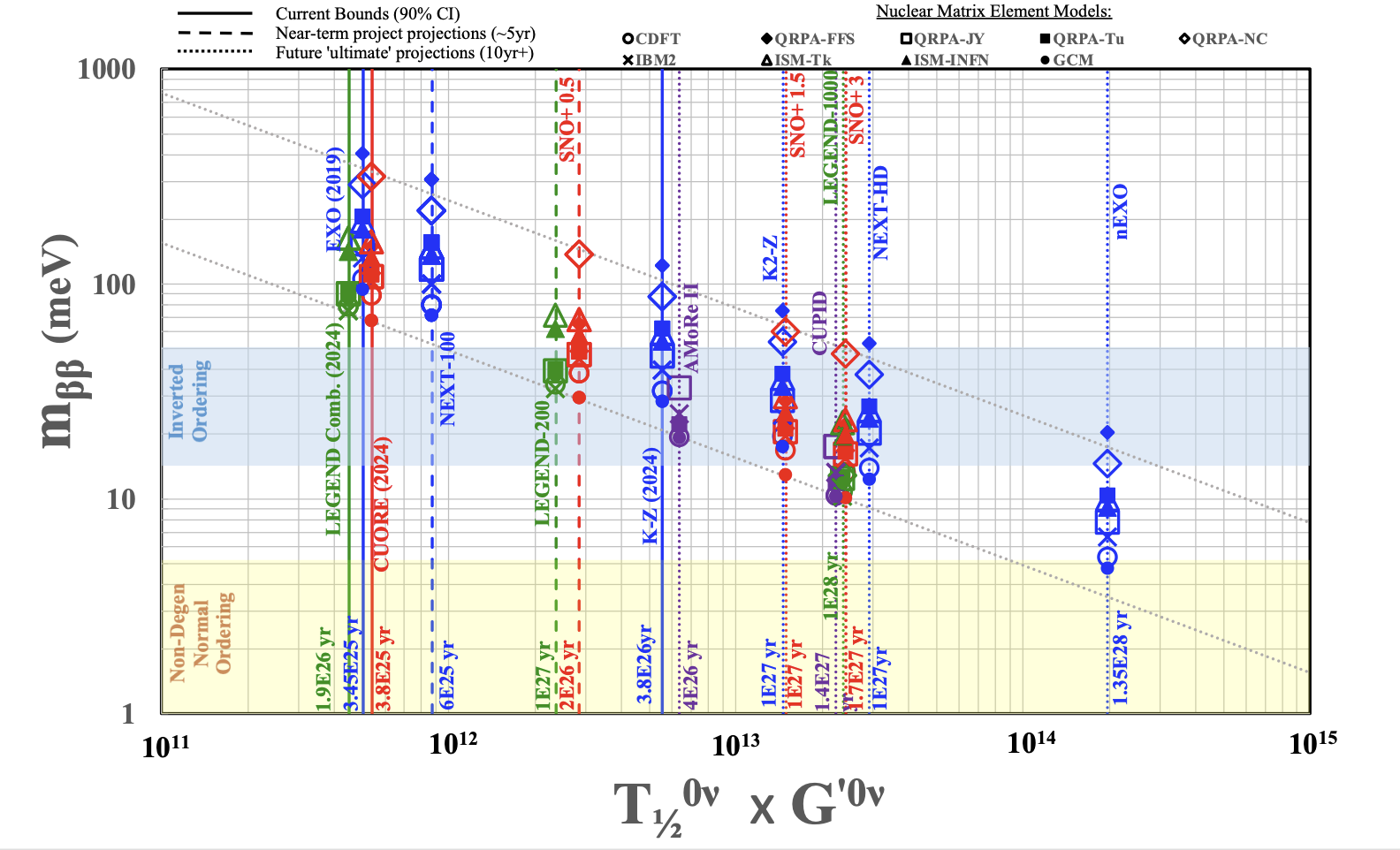}
    \caption{Comparison of $0 \nu \beta \beta$ sensitivities for current~(Section~\ref{sec:topic_lngs_lsm}), near-term and future proposed (Sections~\ref{sec:topic_lngs_lsm} and \ref{sec:nexo}) experiments. Sensitivity is expressed as $m_{\beta\beta}$. Different isotopes are plotted with different colours: $^{76}$Ge, $^{136}$Xe, $^{130}$Te, $^{100}$Mo in green, blue, red and purple, respectively. SNO+ (in red) is shown with increasing amounts of Te (0.5\%, 1.5\% and 3\%). The x-axis is plotted as the expected half-life sensitivity weighted by precisely calculable phase space factors. Different symbols correspond to different nuclear matrix element model calculations.}
    \label{fig:sno+_sensitivity}
\end{figure}

SNO+ will be moving forward with the $0\nu\beta\beta$ phase of the experiment. Tellurium will be added to the detector, starting in 2026, in the form of tellurium-butanediol complexes dissolved in the SNO+ liquid scintillator. The SNO+ $0\nu\beta\beta$ search starts after the current pure scintillator phase has completed its main physics program, and after test batches have fully commissioned the two underground tellurium process plants (purification and synthesis). In the first instance, 0.5\% Te, by weight, will be the target loading. This amounts to 1.33~tonnes of the isotope $^{130}$Te in the detector, a large quantity due to its high natural abundance of 34\%. A proposal has been submitted to the CFI 2025 Innovation Fund competition. If it is successful, it would enable SNO+ to triple the amount of Te loaded in the detector to 1.5\% Te, reaching 4 tonnes of $^{130}$Te deployed for the $0 \nu \beta \beta$ search. \Figure{fig:sno+_sensitivity} illustrates the sensitivity that SNO+ 0.5\% and SNO+ 1.5\% can reach in comparison to other existing, near-future and planned $0 \nu \beta \beta$ experiments.

Tellurium-130 is an important double beta decay isotope that the global $0\nu\beta\beta$ community needs to continue to develop. The reason is that, with 34\% natural abundance, $^{130}$Te does not require costly or logistically difficult isotopic enrichment. To further emphasize this point, the cost per tonne of {\em isotope} to deploy Te in the SNO+ liquid scintillator, including reagents for the purification of telluric acid, butanediol, etc., is 2.1M~CAD/tonne. This is at least an order of magnitude less expensive than any other isotope. 

The feasibility thus exists to scale up a Te-loaded liquid scintillator to eventually reach sensitivity to $m_{\beta\beta}$ in the non-degenerate normal mass ordering range of parameter space. It is for this reason that JUNO and THEIA (Section~\ref{sec:theia}) are also considering a large-scale deployment of Te in their liquid scintillator detectors. SNO+ represents a critical first step in this endeavour. Demonstrating how well this approach works and the control of backgrounds during Te purification and loading are objectives, as well as serving as a competitive $0 \nu \beta \beta$ experiment searching with excellent sensitivity in new parameter space, with a window of opportunity over the next 10 years.

\paragraph{Summary}
\begin{enumerate}
    \item  Great potential for the SNO+ experimental program for tellurium double beta decay at SNOLAB to yield fruit in the next 10-15 years. Strong support from SNOLAB will help make this happen. 
    \item Includes not only searches for $0\nu\beta\beta$ with competitive sensitivity and discovery potential, but also SNO+ Te double beta decay being an important step for the community to keep advancing R\&D in radiopurity and other aspects of background control with $^{130}$Te. 
    \item Complementarity of isotopes and techniques is a pillar of the global $0\nu\beta\beta$ community. If there is to be a discovery of $0\nu\beta\beta$ (or confirmation of one), especially if $m_{\beta\beta}$ is as low as O(1)~meV, it is important for major efforts with the most promising isotopes to be advancing in step.
\end{enumerate}%
\clearpage
\subsection{DEAP-3600 at SNOLAB}
\label{sec:DEAP}

\authorline{Co-Authors: M.~Boulay, C.~Jillings, M.~Lai}

\bigskip 

DM direct detection experiments in liquid argon (LAr) have historically been designed to search for WIMP candidates. Argon, like xenon, is transparent to its own scintillation light, while being easier to purge and scale to multi-tonne experiments. The largest currently running LAr experiment, DEAP-3600, features approximately 3.3 tonnes of atmospheric LAr, monitored by 255 photomultiplier tubes (PMTs) mounted on a stainless steel vessel and coupled to the acrylic vessel. The spherical symmetry is broken at the top of the vessel by the neck, which hosts the cooling coils and allows for refilling and maintenance operations within the acrylic vessel. The stainless steel vessel is submerged in a water tank veto, whose primary purpose is the active rejection of muons and muon-induced events~\cite{DEAPdetectorpaper}.\\

\paragraph{Recent results with DEAP-3600} Analysis of the second fill (Nov 2016 – March 2020) is still ongoing, while the third and final run is scheduled to begin in summer 2025. Thanks to the data acquired in the second fill, the experiment has recently broadened its physics potential, demonstrating how the low background, exposure, and stability of this technology can serve both MeV-scale DM searches and a deeper understanding of argon properties~\cite{DEAPPlanck}\cite{DEAPactivity}. In fact, the experiment has recently measured the $^{39}$Ar half-life to be (302 $\pm$ 8$_{stat}$ $\pm$ 6$_{sys}$) years, significantly longer than the usually accepted values~\cite{DEAPhalflife}. This finding is now under the scrutiny of other experiments seeking to resolve discrepancies with independent measurements. In parallel, the experiment has published a data-driven quenching model for $\alpha$-particle-induced recoils in LAr, initially constrained using $^{222}$Rn, $^{218}$Po, and $^{214}$Rn relative to $^{210}$Po. The nuclear and electronic contributions were then extrapolated down to 10~keV, resulting in the combined quenching model, which is right now used in DEAP-3600~\cite{DEAPquenchingalfa}.\\

Since the multi-scatter search for ultra-heavy DM candidates~\cite{DEAPPlanck}, the experiment has been expanding the DM search from keV-scale nuclear recoils, where WIMPs are the primary candidates, to MeV-scale deposited energy, including candidates such as 5.5~MeV solar axions and inelastic boosted DM. The energy calibrations motivated by these searches have led to the very first hint of $^8$B electron solar neutrinos undergoing absorption in argon. The analysis is currently under internal review and, once confirmed, will allow for an experimental measurement of the cross-section, marking a milestone for the neutrino physics program of the planned next-generation LAr experiments including DUNE, DS20k (see Section \ref{sec:darkside}), and ARGO (see Section \ref{sec:argolite}). 

On the other hand, the second fill has been fundamental in improving the model for background events induced by $\alpha$-particles. Two primary sources have been identified, both of which motivated a Profile Likelihood WIMP search using the full second-fill data, expected to be published this year, as well as hardware upgrades for the upcoming third fill, whose cooling process began in March 2025. One source of events is induced by $\alpha$-particles released by $^{210}$Po within the acrylic flowguides at the bottom of the neck. These induce nuclear recoils in the condensed LAr film on the flowguides, resulting in scintillation light that is first shadowed by the flowguides and detector geometry and then partially reflected at the liquid-gas interface, approximately 551 mm above the detector equator. Meanwhile, degraded $\alpha$-particles released from dust dissolved in LAr contribute additional background, along with the aforementioned neck $\alpha$ events, down to keV-scale deposited energy, making this the dominant background in the WIMP search~\cite{DEAP231dayswimp}.\\

\paragraph{Plans for the Third Fill} One of the main detector upgrades in preparation for the third fill has been the installation of newly custom-developed flowguides. These have been machined in the low-radon machine shop at the University of Alberta, then coated in pyrene-doped polystyrene at Carleton University. These were subsequently shipped to SNOLAB for installation in a sealed can, over-pressurized with research-grade nitrogen gas, to prevent contamination during transport. Pyrene is an extremely slow wavelength shifter, with a decay constant on the order of 100~ns. This will add a slow component to the scintillation light recorded by the PMTs and thereby enable efficient rejection of neck $\alpha$-induced background through pulse-shape discrimination~\cite{DEAPpyrene}.

The other major upgrade is aimed at suppressing the dust $\alpha$ background. A long dust pipe has been inserted down to the bottom of the inner vessel, where more dust is expected to accumulate, through the neck. From here, the LAr will be filtered and returned to the storage dewar, where it will be re-purged before being reinserted into the inner vessel. The pump, purging, and refill processes will be performed in multiple cycles.

The plan is to fill the inner vessel to 3269~kg, similar to the second fill. The weekly analysis, fully automated, including PMT calibration, light yield, and pulse-shape discrimination, will enable comparison of events in the neck and dust $\alpha$ sidebands against the approximately 20 events per week observed in the second-fill dataset for each population. Additionally, a $^{83m}$Kr deployment is planned for summer 2026, with the goal of performing low-energy calibrations and refining the WIMP search using the combined second- and third-fill run. The objective is a zero-background run for DEAP-3600 and the most comprehensive on-site test of the evaluated background models, which will inform the physics potential and design of the future large-scale LAr experiments DS20k (see Section \ref{sec:darkside}) and ARGO (see Section \ref{sec:argolite}).

\bigskip
\paragraph{Summary}

    Summer 2025 is the expected start of the third and final fill in DEAP-3600, addressing fundamental open questions on surface background rejection in tonne-scale argon-filled chambers while expanding the physics case to include non-WIMP searches and neutrino physics.

\subsection{Exploring sub-GeV Dark Matter in Liquid Argon at SNOLAB: DarkSide20k and Darkside-LowMass} 
\label{sec:darkside}
\authorline{Co-Authors: M.~Boulay, C.~Jillings, M.~Lai}

\bigskip 

The Global Argon Dark Matter Collaboration (GADMC) is the result of joint international efforts by DEAP-3600, miniCLEAN, DarkSide-50, and ARDM, bringing together more than 400 scientists across approximately 100 institutions, all focused on DM searches and neutrino physics in LAr.  The only currently running experiment is DEAP-3600 (see Section \ref{sec:DEAP}). This section will discuss the first planned new experiment from GADMC, called DarkSide-20k (DS20k).

\begin{figure}[h]
    \centering
    \includegraphics[width=\linewidth]{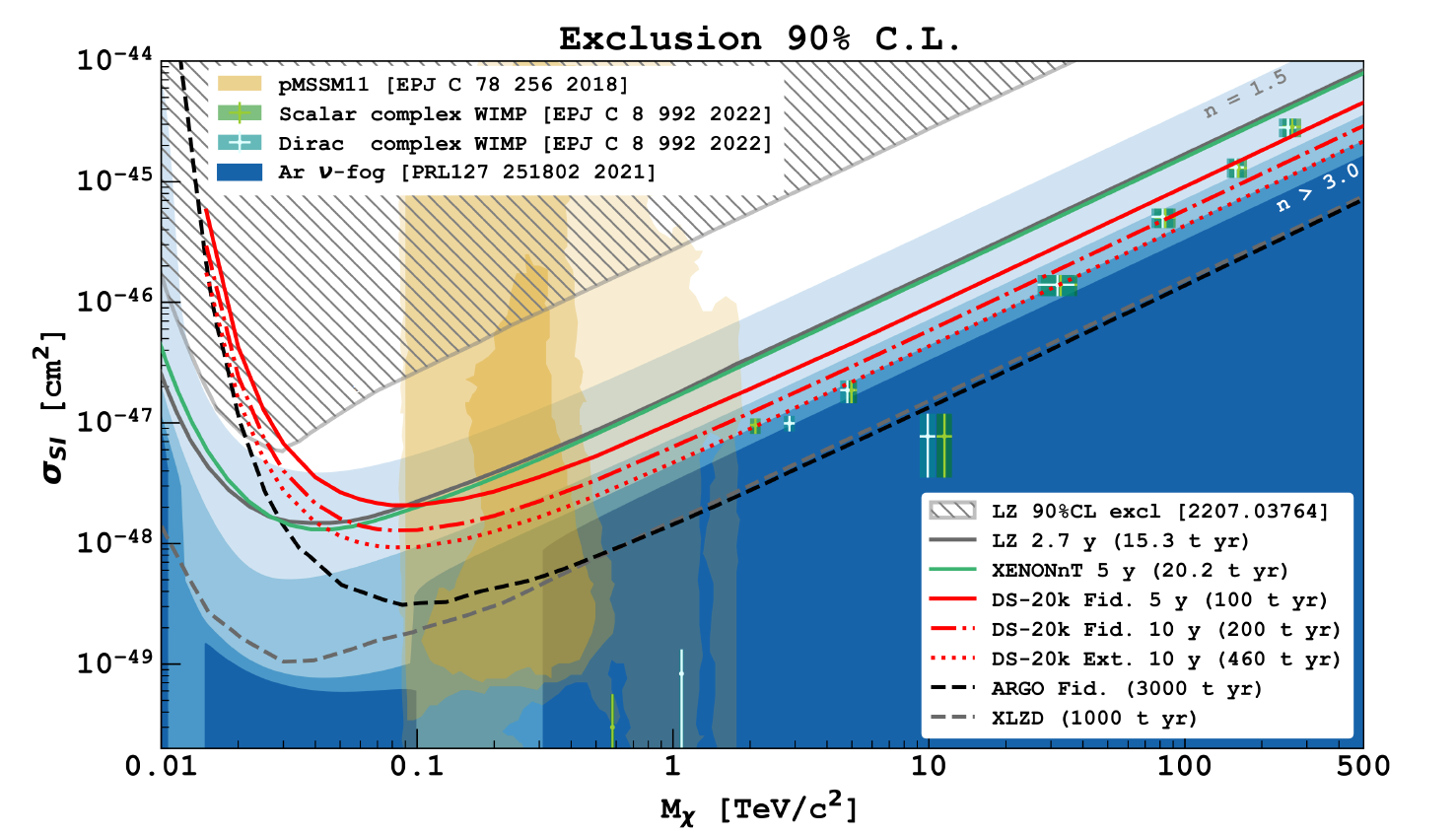}
    \caption{Exclusion limits for spin-independent elastic DM-nucleon scattering for noble liquids in current, upcoming and planned experiments,compared to the main theoretical WIMPs models.}
\label{fig:exclusionlimitshighmass}
\end{figure}

\paragraph{DarkSide-20k} features a 50-tonne, double-phase Time Projection Chamber (TPC) as its main detector, with commissioning expected by early 2028 almost in parallel with DEAP-3600's decomissioning. After ten years of data collection at LNGS, the experiment will scan for Weakly Interacting Massive Particles (WIMPs) down to a DM-nucleon cross-section as low as 5 $\times$ 10$^{-48}$ cm$^2$ for candidates as heavy as 1 TeV/c$^2$. At that time, the experiment will either confirm or exclude WIMPs as the ``missing mass" in our Universe, also verifying or refuting any potential excess observed by XENONnT (see Section \ref{sec:lngs}) and LZ (expected to reach analogous sensitivities using xenon as a target). Achieving sensitivity to such a faint signal also means finally entering the neutrino fog, as shown in Fig.~\ref{fig:exclusionlimitshighmass}.

\begin{figure}[h]
    \centering
\includegraphics[width=0.5\linewidth]{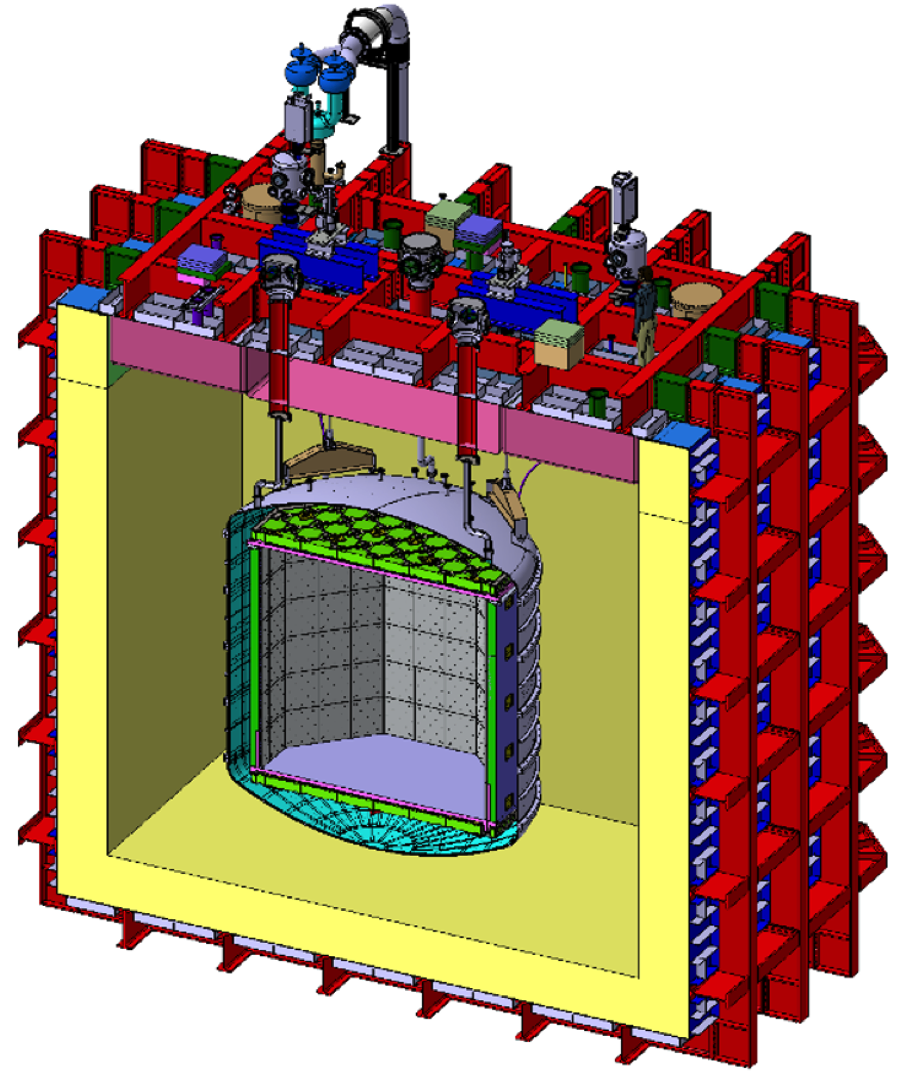}
    \caption{Design of the DarkSide-20k detector.}
    \label{fig:DS20k}
\end{figure}

DS20k includes three nested LAr-filled detectors (see Figure \ref{fig:DS20k}): the 50-tonne double-phase TPC, enclosed within a 72-tonne single-phase detector known as the Inner Veto (IV), which is submerged in a 650-tonne single-phase detector, the Outer Veto (OV). Unlike DEAP-3600, DS20k's design is inspired by DarkSide-50, a 50~kg double-phase TPC that operated at LNGS (see Section \ref{sec:lngs}) until early 2018. Detection in argon chambers, both single- and double-phase, allows for pulse-shape discrimination between $\beta$-particle- and $\gamma$-ray-induced electron recoils (ER) and neutron- or WIMP-induced nuclear recoils (NR) using the scintillation signal (S1), which provides a prompt signal detected with an efficiency of approximately 20\% for NRs. However, double-phase TPCs like DarkSide-50 and DS20k feature an $\mathcal{O}(100)$~V/cm vertical electric potential that drifts most ionization electrons produced by argon de-excitation toward an extraction grid, where they are accelerated by an $\mathcal{O}(1)$~kV/cm potential, generating an amplified electroluminescence signal (S2). This allows for the rejection of multi-site events and refined position reconstruction. The detection efficiency of ionization electrons is approximately 100\%, essentially down to the single electron.\\

\paragraph{DarkSide-LowMass} While standard WIMP searches in DarkSide-50, relying on S1~+~S2 events, could reach an energy threshold of approximately 10~keV$_{NR}$, an S2-only search lowered this threshold to about 0.4~keV$_{NR}$, yielding the most stringent limits for DM candidates at sub-GeV scales~\cite{DarkSide-50lowmass2022}. This breakthrough was a result of the high-quality calibrations, stability, and extremely low background, despite the inability to perform fiducialization along the vertical axis or pulse-shape discrimination when discarding S1 pulses. Consequently, GADMC recently proposed a dedicated low-mass-focused detector as a parallel and complementary project to DS20k: a 1.5-tonne (1-tonne fiducial) double-phase TPC optimized for S2-only DM searches, referred to as DarkSide-LowMass (DSLM) within the argon community \cite{DSLMwhitepaper}. The main detector will sit in approximately 5~tonnes of LAr, serving, analogous to the DS20k IV, as an active veto for neutrons expected to be captured in the acrylic vessel separating the TPC from the LAr bath, resulting in $\gamma$-induced events coincident within the two volumes. The full bath will be housed in a stainless steel vessel submerged in a water tank, serving as an active muon veto, similar to DEAP-3600 (see Section \ref{sec:DEAP}). The primary plan involves placing DSLM's inner detector inside the DEAP-3600 water tank at SNOLAB, leveraging existing infrastructure requested for ARGOLite under the 2025 CFI IF application by GADMC-Canada, as well as Canada's expertise in low-background environments. Key contributors include the low-radon facility at the University of Alberta for acrylic vessel assembly, TRIUMF for DAQ expertise and DS20k efforts, and the cryogenic laboratory at Carleton University.\\

\begin{figure}[h]
    \centering
    \includegraphics[width=0.75\linewidth]{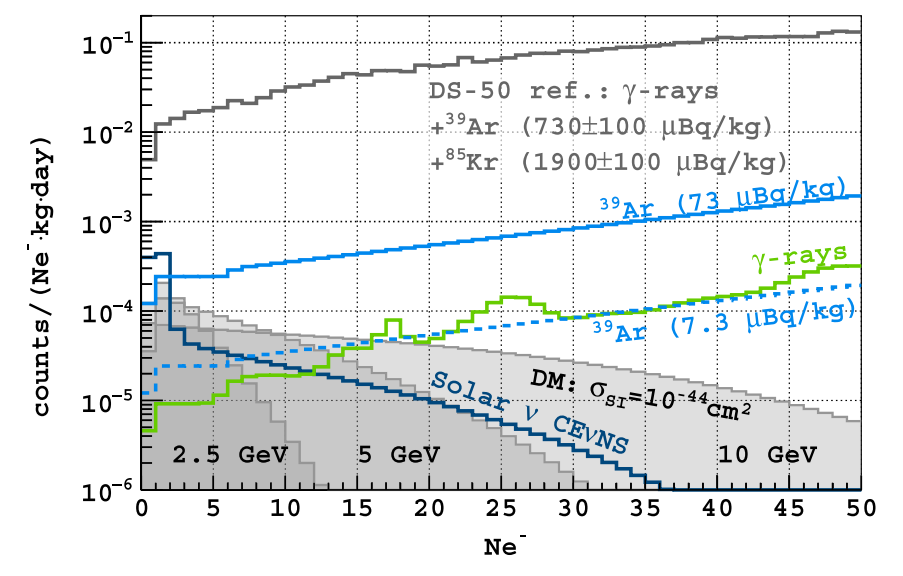}
    \caption{DSLM backgrounds compared to sub-GeV DM candidate energy spectra~\cite{DSLMwhitepaper}.}
    \label{fig:DSLMbackground}
\end{figure}

\paragraph{Backgrounds and Purity} A sensitivity study based purely on state-of-the-art technology shows that the primary irreducible background in DSLM will be solar neutrinos undergoing Coherent Elastic Neutrino-Nucleus Scattering (CE$\nu$NS) at $\mathcal{O}(1)$ ionization electron energies. Any remaining $^{222}$Rn daughters following assembly in the aforementioned low-Rn environment will be further reduced using charcoal traps or molecular sieves, reaching levels comparable to DarkSide-50 and DEAP-3600. Figure \ref{fig:DSLMbackground} presents the background budget, assuming surface backgrounds contribute less than 10\% of $\gamma$-ray-induced events, compared against DarkSide-50's second-fill background, solar neutrino CE$\nu$NS background, WIMP signals for various masses, and the $^{39}$Ar $\beta$-decay background.

Two purification scenarios are considered for the DSLM target, which will consist of UAr extracted from the URANIA plant in Southwest Colorado. Inspired by DarkSide-50, where a 1400-fold reduction in $^{39}$Ar activity was observed in argon extracted from an existing mine and purified at FNAL at 140~g/day, URANIA was built to meet tonne-scale experimental needs and is expected to begin commissioning by summer 2025 with an extraction rate of approximately 250~kg/day. The UAr will then be processed at ARIA using the 350-meter-tall Seruci-1 cryogenic isotopic distillation column to reduce $^{39}$Ar down to $\mathcal{O}(10)$~$\mu$Bq/kg, compared to $\mathcal{O}(1)$~mBq/kg achieved in DarkSide-50~\cite{UArpath}. However, achieving this level of purification requires a 10~kg/day throughput at ARIA, which is feasible for DSLM but unrealistic for DS20k, where ARIA will be used only for standard chemical distillation. Including cosmogenic activation of $^{39}$Ar during surface storage and transport to and from North America, DSLM's final $^{39}$Ar content is expected to be approximately 73~$\mu$Bq/kg after one pass through ARIA and 7.3~$\mu$Bq/kg after two~\cite{DSLMwhitepaper}.

\begin{figure}[h]
    \centering
    \includegraphics[width=\linewidth]{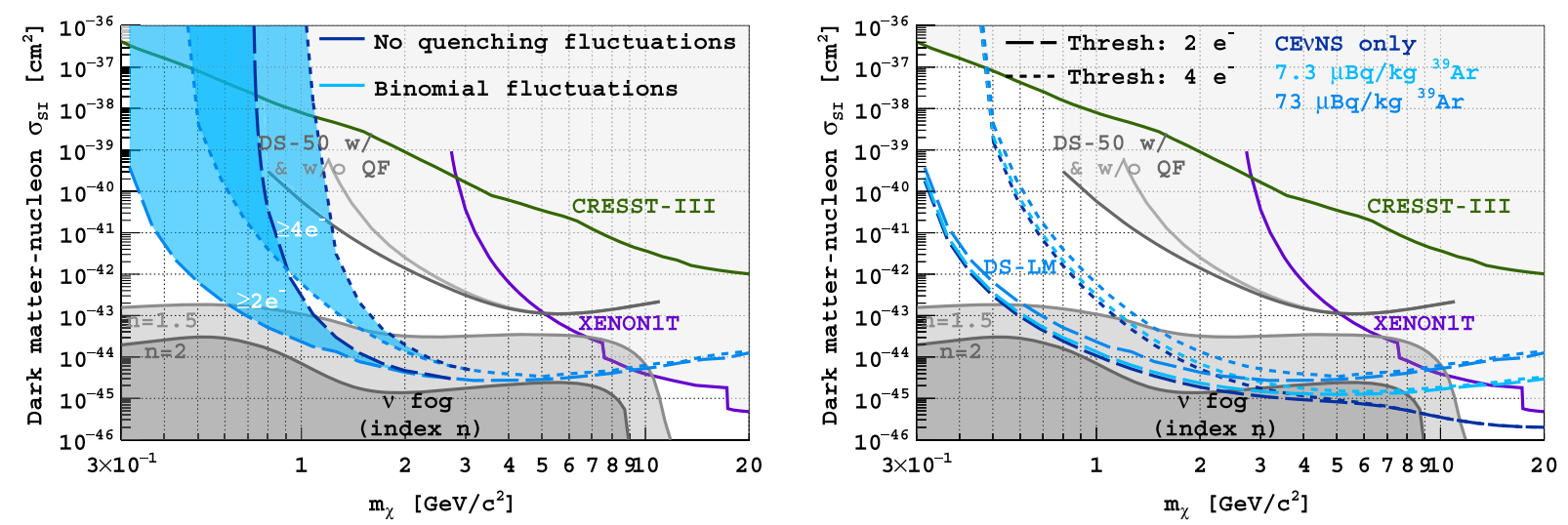}
    \caption{Exclusion limits at 90\% C.L. for DSLM~\cite{DSLMwhitepaper}}
    \label{fig:DSLMsensitivity}
\end{figure}

An additional background at [1-4]~ionization electrons was identified during DarkSide-50's UAr run, linked to impurities dissolved in LAr, as its rate increased when the getter was accidentally switched off. Preliminary studies indicate the presence of uncorrelated background components, including signals at approximately 40~mHz with decay times of 8~ms and 48~ms, and a 90~mHz component with a decay time of 15~ms appearing only when the getter is off. The latter is possibly due to trapped electrons released shortly after switching it off, which then enter another acquisition window/event. Ongoing R\&D efforts aim to pinpoint the source of these backgrounds by investigating correlations with target purity, TPC electric fields, ionization yield, and inner detector materials. The exclusion limits for two different analysis threshold (at 2 and 4 ionization electrons) and two UAr purity levels after isotopic distillation are shown in Fig.~\ref{fig:DSLMsensitivity}.

\paragraph{Dopants} In addition to background suppression, DSLM is exploring new target materials, aiming to lower its analysis threshold by adding a few percent of molecules that undergo ionization from argon scintillation light. These are so-called ``photosensitive dopants.'' Xenon-doped argon has been the primary focus, where, alongside the usual VUV light from argon de-excitation at 128~nm, additional emissions at 150~nm from Ar-Xe excited dimers and 178~nm from xenon dimers contribute~\cite{xenondopedSegreto}. At $\mathcal{O}(10)$~ppm xenon doping, the scintillation light shifts from 127~nm to approximately 174~nm, enabling in-situ wavelength shifting and eliminating the need for TPB, a suspected source of spurious electron backgrounds. However, while the scintillation yield increases by approximately 20\% at 100~ppm xenon doping, this comes at the cost of a substantial loss in pulse-shape discrimination efficiency due to xenon triplet states decaying after $\mathcal{O}(10)$~ns~\cite{Galbiati:2020eup,xenondopedDUNE}.\\

Further research is ongoing for hydrogenous dopants such as allene, TMG, and TMA, which ionize at 7.5–9.5 eV, converting inefficiently detected argon scintillation light into additional S2 signals~\cite{Andersondopedargon}. High doping levels may also enhance sensitivity to WIMP masses 40 times lower than conventional limits, expanding the range of testable models in DSLM. Additional measurements are planned to calibrate doped-argon ionization response down to sub-keV scales, leveraging existing LAr setups. Further background rejection studies for tonne-scale doped argon detectors are scheduled at Queen's University starting fall 2025, within GADMC efforts and in synergy with SBC (see Section \ref{sec:sbc_summary}) and DUNE~\cite{xenondopedSBC} \cite{xenondopedDUNE}.

\bigskip
\paragraph{Summary}

    DM searches in DarkSide-LM are planned for 2032-2038, investigating candidates at the sub-GeV scale, first with pure argon and later with the optimal photosensitive dopant.

\subsection{The Global Argon Dark Matter Collaboration at SNOLAB: R\&D with ARGOlite}
\label{sec:argolite}
\authorline{Co-Authors: M.~Boulay, C.~Jillings, M.~Lai}

\bigskip 

The final milestone of GADMC (Global Argon Dark Matter Collaboration, introduced in Section \ref{sec:darkside}) will be ARGO. A 400-tonne (300-tonne fiducial) detector, its  ten years of data collection will allow for the complete exclusion of WIMPs and their detectability in argon, and, together with XLZD (Section~\ref{sec:xlzd}), in noble liquid detectors. Besides providing the ultimate answer on WIMPs in argon, ARGO will showcase a broader physics program, including non-WIMP searches and ultra-heavy DM candidates, thanks to its large cross-sectional area and the potentially instrumental background-free search for solar neutrinos and neutrinos from supernovae~\cite{DS20ksupernova}. These efforts will build upon ongoing analyses and sensitivity studies in DEAP-3600 (see Section \ref{sec:DEAP}) and DS20k (see Section \ref{sec:darkside}).

The main detector will be either a single-phase chamber or a double-phase TPC. That design will depend on the initial results from DS20k in the early 2030s, specifically in relation to the actual gain in position reconstruction and background rejection, which has historically been attributed to the ionization signal in the gas pocket. The target will be stored in an acrylic vessel submerged in a LAr vessel, all surrounded by an outer cryostat, likely consisting of a vacuum cryostat along with a water tank to reduce external neutron contamination. Two main algorithms for position reconstruction and clustering are under development: one fully based on time information and the other based on hit-pattern mapping along the photosensors. The first outcome from these simulations will be constraints on the allowable $^{210}$Po surface contamination, which will inform construction and integration procedure requirements.\\

In parallel with the simulations, in the early 2030s, the collaboration plans to assemble a multipurpose low-background prototype at SNOLAB, named \textbf{ARGOLite}. ARGOLite will consist of one tonne of ultrapure argon extracted from URANIA~\cite{uraniaproceeding}, to be installed in the DEAP-3600 water tank after its decommissioning. ARGOLite will allow for the study of surface $\alpha$ backgrounds as low as 10~mBq/m$^2$, with a background leakage in the WIMP ROI as low as 10$^{-7}$, while maintaining an electron-recoil/nuclear-recoil discrimination power of approximately 10$^{-9}$. ARGOLite will also enable the measurement of $^{42}$Ar and $^{85}$Kr in underground argon (UAr) from URANIA, confirming the expected contamination in the DS20k target.\\

A major application for ARGOLite will be the deployment of pixelated silicon photon avalanche diode (SPAD) arrays, developed by the University of Sherbrooke, which are expected to fully cover the 250~m$^2$ photosensitive area in ARGO~\cite{digitalsipms}. ARGOLite will feature 2~m$^2$ of pixelated SPADs. These SPADs incorporate CMOS readout electronics on the back of the pixels, significantly reducing electronic noise and enabling selection and ``shutting down'' of individual SPADs exhibiting higher dark noise rates. Furthermore, additional active quenching can be selectively applied to minimize afterpulses, while isolation trenches help reduce optical cross-talk between SPADs. Currently, two electronic systems are being compared. The full-photon-to-digital converter (PDC) system, developed at the University of Sherbrooke, features each SPAD bonded to its own readout circuit in a 3D configuration, with digital information subsequently grouped into ``pixels''~\cite{ditialsipmPDC}. In contrast, the ATARI ``2.5D'' system  -- developed at LNGS (see Section \ref{sec:lngs}) and the University of L'Aquila -- first aggregates SPADs into pixels, which are then bonded to their respective readout circuits. While the ATARI approach is simpler, it sacrifices some of the flexibility provided by pixelated technology. As the two systems are compared, TRIUMF will oversee integration and readout, leveraging expertise developed within DS20k.

Another primary development for ARGO will be the design of an optimized isotopic distillation column to reduce $^{39}$Ar levels in the UAr target, whose $\beta$-decay events contribute to dead time, even though they can still be rejected through pulse-shape discrimination~\cite{DEAPPSD}. Unlike the purification goals outlined for ARIA\cite{ARIA2}, ARGO will require medium purity, which scales with column height, but high purification throughput, which scales with diameter. Based on earlier studies, a 200–300~m tall column can be optimized to approximately 1~m in diameter, corresponding to a purification throughput of $\mathcal{O}(100)$~kg/day. Currently, GADMC-Canada has proposed a 4-inch-diameter, 20-m-tall distillation column to be installed at SNOLAB as a prototype, aimed at finalizing and demonstrating the necessary design for ARGOLite and the ARGO distillation column. Ideally, this system would be installed at Shaft~8 at SNOLAB, in collaboration with the xenon community working on Xe-enrichment for $0\nu\beta\beta$ (see Section~\ref{sec:nexo}).

\paragraph{Summary}
\bigskip
\label{sec:gadmc_summary}

\begin{enumerate}
\item The GADMC will develop the necessary hardware for ARGO over the next decade, ultimately aiming to discover or exclude WIMPs interacting with argon. 
    \item 2025-2035: New calibrations at local institutions to extend the calibration of pure argon as well as doped-argon response down to sub-keV-scale energy recoils.
    \item 2028-2031: ARGOLite at SNOLAB, serving as a multipurpose prototype for ARGO and a potential inner veto for DarkSide-LM.
     \item Late 2030s: Construction of ARGO, preferably at SNOLAB.

     \item We anticipate the following major upgrades at SNOLAB:
\begin{itemize}
    \item Radon-reduced air in the Cube Hall to ensure efficient commissioning of ARGOLite and DarkSide-LM, and later ARGO, by maintaining the lowest achievable surface background levels.
    \item Upgrades to the existing chemistry facility, including a dedicated setup for radon removal and noble liquid target purification before insertion into DSLM and later ARGO.
    \item Installation of an isotopic cryogenic distillation column, possibly at Shaft 8, to remove $^{39}$Ar from UAr at SNOLAB at a rate of O(100) kg/day, ensuring sufficient purification for ARGO.
    \item Storage and logistics for 400 tonnes of low-radioactivity argon for ARGO's main detector.
\end{itemize}
\end{enumerate}

\subsection{nEXO 2.0: A Next-Generation Neutrinoless Double Beta Decay Experiment with Liquid Xenon}
\label{sec:nexo}

\authorline{Co-Authors: T. Brunner, E. Caden}

\bigskip
nEXO is a next-generation $0\nu\beta\beta$ decay experiment searching for this decay in the isotope $^{136}$Xe in a 5~tonne single-phase liquid Xe (LXe) TPC. The experiment has been developed by an international collaboration of more than 200 scientists at 39 institutions in 9 countries. Following the recent decision by the US Department of Energy Nuclear Physics Division to advance LEGEND-1000 to CD-1 (Critical Decision 1) while maintaining an R\&D program for nEXO, the Canadian team has taken leadership. The collaboration has developed a phased approach called ``nEXO 2.0'' with the goal of a timely construction start at SNOLAB's Cryopit. This phased approach anticipates construction start ideally in 2026 with a commissioning of the detector in 2032 or 2033. The detector will be based off the positively reviewed nEXO design, but in the initial phase it will be filled and operated with 5~tonnes of natural xenon. Following operation of about 2 years, the natural xenon will be replaced with xenon enriched to 90\% in the $\beta\beta$-decaying isotope $^{136}$Xe and operated for $\sim 8$ years to reach a sensitivity beyond $10^{28}$ years.

This material presented at the FPW outlined the scientific motivation and strategic opportunity for Canada to lead a xenon-136-based search for $0\nu\beta\beta$ at SNOLAB. The discovery of $0\nu\beta\beta$ would have profound implications for fundamental physics, including insights into the quantum nature of neutrinos, the mechanism of matter creation, and the origin of the universe’s baryon asymmetry. Since $0\nu\beta\beta$ is forbidden in the Standard Model, its observation would signal new physics. The international physics community has long recognized the significance of this search, and both the Canadian and U.S. long-range plans have identified $0\nu\beta\beta$ as a strategic priority.

The presentation reviewed the current status of the nEXO experiment, a LXe TPC with a 5-tonne active mass, which has undergone multiple successful technical and scientific reviews. Despite recent DOE decisions to prioritize funding for LEGEND-1000 (see Section \ref{sec:lngs}) in the near term, the nEXO concept was found to demonstrate high potential and technical readiness. Canada has played a central role in the development of nEXO over more than a decade and possesses relevant expertise in liquid noble gas detector technologies, as well as access to SNOLAB --- a suitable deep-underground facility for hosting the experiment.

The proposed path forward includes forming a renewed collaboration around a Xe-136 experiment at SNOLAB (referred to as “nEXO 2.0”), aiming for a sensitivity of $10^{28}$ years within ten years of data taking. A phased approach to detector construction was presented, offering flexibility in the initial xenon enrichment and enabling potential upgrades over time. Infrastructure layout studies demonstrate feasibility within the SNOLAB Cryopit, and existing CFI IF 2020 and IF 2023 infrastructure funding awards are available to support immediate R\&D and detector readiness. A timeline was outlined with the goal of beginning construction by 2026.
\bigskip
\paragraph{Summary}
\begin{enumerate}
    \item Discussions at FPW centered on the implications of the DOE decision and the resulting opportunity for Canada to step into a leadership role. There was interest in building a path forward that is technically feasible, internationally collaborative, and competitive in scope and schedule with other major global $0\nu\beta\beta$ initiatives, such as LEGEND-1000 (see Section \ref{sec:lngs}) and PandaX (see Section \ref{sec:topic_CJPL}). 
    \item Participants also discussed the importance of resolving procurement and logistical challenges associated with enriched Xe-136, and emphasized the urgency of maintaining project momentum. 
    \item A follow-up workshop was proposed to evaluate detector technologies and develop a roadmap for a competitive xenon-based $0\nu\beta\beta$ program at SNOLAB.
\end{enumerate}%
\subsection{NEXT: Neutrino Experiment with a Xenon TPC}
\label{sec:topic_next}

\authorline{Co-Authors: K. Mistry (on behalf of the NEXT Collaboration)}

\bigskip

The Neutrino Experiment with a Xenon TPC (NEXT) is an experiment searching for $0\nu\beta\beta$ with high-pressure xenon gas enriched in $^{136}$Xe. This detector offers several benefits, including 0.5-1\% energy resolution at 2.5~MeV through electroluminescent amplification of ionization charge, powerful topological separation of signal (two electron signals) vs. background (single electron signals), scalability, and the potential for tagging the Ba$^{2+}$ daughter from the decay enabling a background-free experiment.

The program is phased, with the current experiment NEXT-100~\cite{next100detector,next100CDR}, currently under operation. Initial data from NEXT-100 has achieved electron drift lifetimes of the order of tens of ms, energy resolution of 5\% at 40~keV with $^{83\textnormal{m}}$Kr calibration, and reconstruction of 30~keV x-rays. NEXT-100 is currently taking calibration with a high-energy $^{232}$Th source (2.6~MeV gammas). 

NEXT-100 is expected to operate for the next 5 years with potential upgrades to the detector during this period. Following this, NEXT is planning a tonne-scale multi-module program~\cite{NEXT1t} beginning construction in 2032, with target sensitivities to the half-life beyond $10^{27}$ yr at 90\% confidence level. This program includes NEXT-HD and NEXT-BOLD. For the first module, NEXT-HD, an extensive R\&D program is underway. This includes the implementation of a fibre-barrel, new ASICs, metalenses, and a dense tracking plane. The second module, NEXT-BOLD, includes barium tagging~\cite{Jones:2016qiq,McDonald:2017izm}. By tagging the barium ion, in-coincidence with the two-electron signal at 2.5~MeV this will enable an essentially background free experiment and boost sensitivities of the half-life from $10^{27}$~yr toward $10^{28}$~yr after a 10~tonne-year exposure time. Excellent progress has been made to realize the barium tagging technology including recent demonstrations of RF carpet transport~\cite{NEXTRFCarpet} and time-resolved sensors~\cite{Aranburu2025}. 

\paragraph{Summary}
\begin{enumerate}
    \item The Laboratorio Subterráneo de Canfranc (LSC) is enthusiastic to host a NEXT-tonne scale module, but this has not been decided yet. SNOLAB is an ideal candidate for hosting such a module due to its depth and facilities. Backgrounds from $^{136}$Xe neutron activation to $^{137}$Xe would be highly suppressed to negligible levels. Clean room facilities are world-leading and is ideal for hosting a low-background double beta decay experiment. The Cryopit area would be the perfect location since it will have ample space for a water shield of 3~m surrounding the detector which will be 2.6~m in size and operated at 15~bar pressure. 
    \item There is clear interest and possibility of significant financial backing from Canada for a discovery-class $0\nu\beta\beta$ experiment to be hosted at SNOLAB with xenon based-technology. This builds on the existing investments from Canada in nEXO. The details of exactly what technology would be used, which institutions and collaborations would contribute to this is currently in significant flux. This also includes the possibility of merging efforts from a $0\nu\beta\beta$ program and DM program within XLZD (see Section \ref{sec:xlzd}), for example. 
    \item There are ongoing discussions among institutions from across the portfolio of xenon-based technologies ($0\nu\beta\beta$ and DM) and several workshops are being organized with one in Heidelberg on 26th May (the Third International Summit on the Future of Neutrinoless Double Beta Decay~\footnote{\url{https://indico.ph.tum.de/event/7802/}}) and another to be hosted in Montreal in November 2025 (Neutrinoless double beta decay search in Xe - next-generation experiment workshop~\footnote{\url{https://nyx.physics.mcgill.ca/event/538/}}). 
    \item A key discussion point is exactly what would be the target sensitivity to achieve for double beta decay. It was mentioned that this experiment should achieve meV mass sensitivity which translates to $10^{30}$~yr half-life. At present, even if the experiment is completely background-free, to achieve such sensitivities requires huge masses of hundreds of tonnes or even a kilotonne of xenon. It is not clear if such an amount of xenon could be obtained with current methods, and even further whether we can obtain this amount enriched in $^{136}$Xe. This necessitates global collaboration. Otherwise, the $10^{28}$~yr half-sensitivity seems like the limit of what could be achieved. 
    \item A tonne-scale detector sited at SNOLAB would require expansion of personnel (engineers, scientists, postdocs and students) at SNOLAB to support its construction. In the NEXT case, experts in high pressure gas would be required. New facilities to ensure xenon recovery and detector construction may be required. Additional support for efficient radioactivity screening will also be required. 
    
\end{enumerate}

\subsection{XLZD: A Liquid Xenon Underground Rare Event Observatory}
\label{sec:xlzd}

\authorline{Co-Authors: W. H. Lippincott, on behalf of the XLZD Collaboration}

\bigskip

The XLZD underground rare event observatory based on liquid xenon (LXe) technology will address some of the most important open questions in fundamental physics and cosmology: the nature of DM, which drives the formation of structures in the universe such as galaxies and clusters, and the fundamental nature of neutrinos, which is closely tied to the puzzling matter-antimatter asymmetry in the universe. XLZD will conduct highly sensitive measurements with its quiet, massive detector, offering an unrivaled low energy threshold and background level required to tackle these mysteries. Additionally, XLZD will perform precision measurements of solar neutrinos, search for solar axions, and watch for neutrinos from supernovae in our cosmic neighborhood. With its rich scientific program, XLZD will thus be a true multi-purpose observatory in astroparticle physics, poised to make a global impact.	

In its nominal design~\cite{XLZD:2024nsu}, XLZD is an observatory with a target mass of 60~tonnes of cryogenic LXe in its central detector. Xenon is an ideal medium for detecting ultra-rare particle interactions. Should the xenon market and funding situation permit, an alternative design would include 80~tonnes of xenon target. It will operate as a dual-phase time projection chamber (TPC). This technology revolutionized  direct DM searches about 20 years ago with its highly scalable detection principle and has since led the field. XLZD is being pushed forward by the leading teams in this area: XENON and LZ. These collaborations currently operate the largest LX2 detectors, with target masses of 5.9~tonnes and 7.0~tonnes, respectively. The DARWIN collaboration is conducting R\&D towards a multi-ton LXe detector~\cite{LZ:2024zvo,XENON:2025vwd,DARWIN:2016hyl}. More details about the proposed design and science goals of XLZD can be found in the XLZD Design Book~\cite{XLZD:2024nsu}. 		

The site for XLZD has not yet been chosen, but four international underground facilities have been selected to form a short list based on technical and scientific requirements for the experiment, including depth of the laboratory, available space, and considerations of fabrication and operation. These sites include SNOLAB; the Boulby Stage 2 expansion in the UK; LNGS in Italy (see Section \ref{sec:lngs}); and SURF in the USA. The selection of the DUL is an important major milestone; the process is under preparation. The decision is expected to be made in 2026. 

Design studies are ongoing and it is expected that these will lead to a Technical Design Report (TDR) before beginning production and installation by the end of the decade. The start of underground construction depends on the selected host laboratory, and because many items will require underground fabrication, early occupancy will be important to maintain a schedule. We anticipate the commissioning of the entire observatory (i.e., with the nominal LXe target) and the start of the science operation phase by the mid 2030s. The overall timeline crucially depends on the Xe procurement activities, which are closely related to the availability of funding: the global Xe market can easily deliver the required amount of gas to XLZD; however, a timely start of the procurement and the negotiation of long-term contracts with a few key suppliers is mandatory. An early science phase with a detector of reduced target mass is possible starting with $\sim$45~t of Xe gas (procurement of 25~t with an additional 20~t from XENON and LZ), provided that the selected DUL is ready for beneficial occupancy.		

\begin{figure}[h!]
    \centering
    \includegraphics[width=0.85\linewidth]{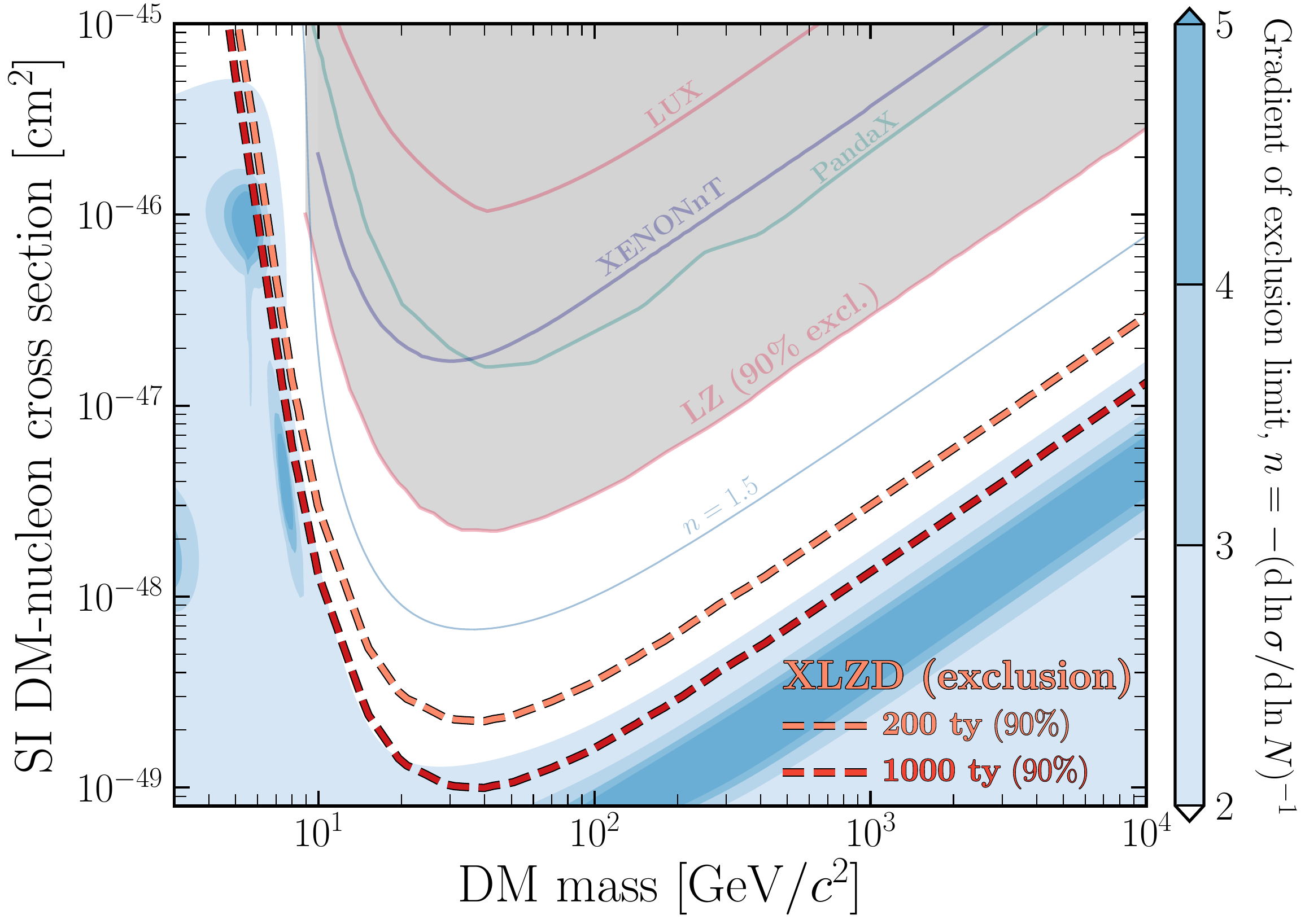}
    \caption{Projected 90\% C.L. upper limits on the spin-independent WIMP-nucleon cross section for 200 and 1000 tonne$\cdot$year (t$\cdot$y) exposures of the XLZD detector, along with current upper limits~\cite{LZ:2024zvo, XENON:2025vwd,PandaX:2024qfu,LUX:2015abn}. The blue shaded regions illustrate the irreducible background from neutrinos (neutrino fog).}
    \label{fig:XLZD_DM}
\end{figure}

\begin{figure}[t!]
    \centering
    \includegraphics[width=0.85\linewidth]{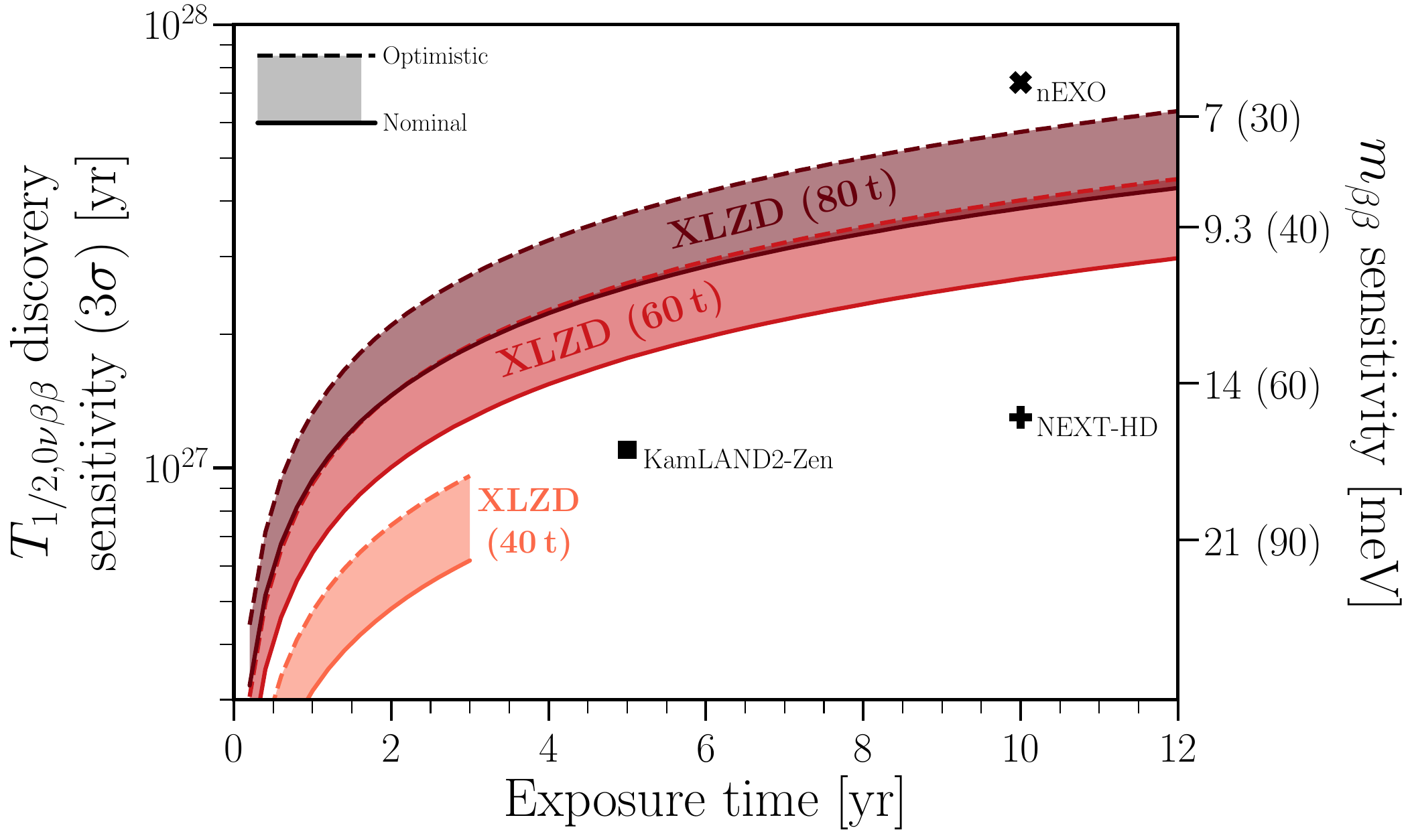}
    \caption{Projected evidence sensitivity at 3$\sigma$ statistical significance to $0\nu\beta\beta$ of $^{136}$Xe as a function of exposure time for the two XLZD target mass scenarios: 60t and 80t, and
an interim 40~t configuration. For each target mass, the band represents a range of detector performance
parameters and background assumptions between the nominal (lower limit) and optimistic (upper limit)
scenarios~\cite{XLZD:2024nsu,XLZD:2024pdv}. The right axis shows the sensitivity in terms of the effective Majorana mass $m_{\beta\beta}$, assuming
a nuclear matrix element of $M^{0\nu}_{136Xe} =1.11$ or 4.77 (in brackets).}
    \label{fig:XLZD_DBD}
\end{figure}

The XLZD observatory is expected to be operated for at least 15 years to fully exploit its extraordinary scientific potential which requires collecting a significant low-background exposure.	Projections of the observatory's sensitivity for DM and $0\nu\beta\beta$ (of the ${}^{136}$Xe component of natural xenon) are shown in Figs.~\ref{fig:XLZD_DM} and \ref{fig:XLZD_DBD}, respectively.	Studies by the collaboration suggest that with 3~years of exposure, XLZD will become the most sensitive $0\nu\beta\beta$ experiment with a $^{136}$Xe target~\cite{Agostini:2022zub,nEXO:2021ujk,NEXT:2020amj}

\paragraph{Summary}

The XLZD low-background observatory will:	

\begin{enumerate}
\item Investigate the nature of DM and neutrinos, and will study astrophysical objects (the Sun, galactic supernovae) via neutrinos:
\begin{enumerate}
    \item Probe DM into the irreducible background of the atmospheric neutrino fog at masses above 10~GeV~\cite{XLZD:2024nsu}
    \item Explore the $0\nu\beta\beta$ mode of 136Xe with a half-life sensitivity approaching $10^{28}$~years~\cite{XLZD:2024pdv}
    \item Measure the solar pp-neutrino flux with unprecedented precision~\cite{Aalbers:2022dzr}
    \end{enumerate}
\item Use a highly sensitive dual-phase LXe TPC with a 60~t active target mass, where the background will be dominated by neutrino interactions. Should the xenon market and funding situation permit, an alternative design would include 80~t of xenon target.
\item Operate as an international flagship experiment in the 2030s and into the 2040s.
\item Carry out an appropriate site selection process with a plan to make a decision in 2026. 

\end{enumerate}%
\subsection{Theia: A Hybrid Neutrino Detector}
\label{sec:theia}

\authorline{Co-Authors: M. Wurm, on behalf of the Theia proto-collaboration}

\bigskip

We propose that SNOLAB could be an ideal site to host ``Theia''. Theia is a novel, ``hybrid'' neutrino detector able to observe both the directional photons of the Cherenkov light signature and isotropic photons from the much brighter scintillation signal. The combined information facilitates a rich program of cutting-edge, interdisciplinary science. Theia's hybrid technology will interrogate a broad range of neutrino physics aligned with community priorities, with world-leading sensitivities and complementarity to other large-scale neutrino experiments. 

Bright scintillation light provides the necessary energy resolution for low-energy neutrino physics, including measurements of the CNO solar neutrinos, antineutrinos from the crust of the Earth (geo-neutrinos), and eventually a search for $0\nu\beta\beta$ with sensitivity into the normal neutrino mass-ordering hierarchy. The ability to detect and identify Cherenkov photons will provide directional sensitivity, critical for identification of signatures such as solar neutrinos. Measuring the ratio of Cherenkov and scintillation light will allow Theia to reject backgrounds to signals such as the Diffuse Supernova Relic neutrinos (DSNB). Theia would enable a unique multi-messenger astronomical measurement for supernova burst neutrinos. This is can do by detecting anti-electron neutrinos coupled with directional sensitivity from electron scattering.

\paragraph{Summary}
\begin{itemize}
\item Table~\ref{tab:theia} provides an overview of key numbers of the Theia physics program. The full breadth of the this program is explored in the Theia white paper~\cite{Theia:2019non}.
\item While scalable, the astrophysical program of Theia requires a minimum of 25,000 tons of hybrid scintillator and a new detector cavern of corresponding dimensions. SNOLAB --- given its exceptional rock shielding and making use of the excellent infrastructure and experience for low-background physics in general and the SNO(+) detector programs in particular --- would be the ideal laboratory to facilitate both the astrophysical neutrino and double-beta decay programs of Theia. 
\end{itemize}

\begin{table*}[!h]
\centering
\caption{Theia physics reach.  Exposure is listed in terms of the fiducial volume assumed for each analysis. For $0\nu\beta\beta$ the target mass assumed is the mass of the candidate isotope within the fiducial volume.
\label{tab:theia}}
\begin{tabular}{l l l} 
\hline\noalign{\smallskip}
 {\bf Primary Physics Goal} & {\bf  Reach} & {\bf  Exposure} \\
\noalign{\smallskip}\hline\noalign{\smallskip}
    Long-baseline oscillations 	& 	$> 5 \sigma$ for 30\% of $\delta_{CP}$ values	&	524 kt-MW-yr \\
    Galactic Supernova (10\,kpc) &	$<2^\circ$ pointing accuracy	& 25-kt detector \\
 								& 5,000 events  &\\
   DSNB 	&	$ 5 \sigma$ discovery	&	125 kton-yr \\
    CNO neutrino flux	&	$< $ 10\%	& 62.5 kton-yr	\\
     Reactor neutrino detection	&	2000 events	&	100 kton-yr \\
     Geo neutrino detection 	&	2650 events	& 100 kton-yr	\\
     $0\nu\beta\beta$               	&    	T$_{1/2} > 1.1\times10^{28}$~yr	 &     211 ton-yr $^{130}$Te	\\
     Nucleon decay $p\rightarrow \overline{\nu}K^{+}$ 	&	$T>3.80\times10^{34}$~yr (90\% CL)	& 800 kton-yr	\\
\noalign{\smallskip}\hline
\end{tabular}
\end{table*}%

\newpage
\section{Perspectives on the Future}

Multiple workshop participants provided perspectives that go beyond the scope of any one defined project and could provide pathways to future collaborations, concepts, or programs. Three such perspectives are presented here, the first two on $0 \nu \beta \beta$ and the third on the cosmic neutrino background.

\hyphenation{CHOCOLATE}

\subsection{The Way Forward for Neutrinoless Double Beta Decay}
\label{sec:future_0vbb}

\authorline{Author: D. Sinclair}

\bigskip

The search for $0\nu\beta\beta$ is a very high priority for the astroparticle physics community. In this note I will argue that the focus of new facilities for this search should be aimed at reaching the bottom of the normal hierarchy (or normal ordering, NO) band. I will also suggest ways in which current liquid xenon experiments might achieve the required energy and spatial resolutions to achieve the required background levels. The realities of the world xenon production capabilities imply that this project should be done as a global initiative and the best time to build such collaborative effort is now when everyone can contribute to the development of the design.

\subsubsection{Introduction}

The discovery by SNO and Super-Kamiokande that neutrinos oscillate and thus possess mass, and the discovery that neutrino mass states have large amplitude mixing of the lepton number states, are twin observations that greatly increased the interest and urgency for searching for $0\nu\beta\beta$. To date, the searches have been sensitive to only the highest values of the lightest mass state, $m_1$, allowed by the constraints of cosmology and the KATRIN experiment~\cite{KATRIN:2025}. 

To cover the complete space of possible values for $m_1$ one needs to extend the searches to a value of $m_{\beta\beta}$ of about 1~meV, thus covering the complete band for the NO. Such a search would become even more compelling if, in a few years, the long baseline neutrino experiments confirm the present indications of NO. To complete such a search using liquid xenon, one needs to reduce beta-gamma backgrounds by a factor of about 100 compared with nEXO~\cite{nEXO:2018ylp} and to make significant improvements in energy resolution to control the background from the two neutrino double beta decay. In this contribution I will discuss techniques that might allow such improvements. The intent of this contribution is not to define a specific detector, but to stimulate interest in new ideas for the detection of $0\nu\beta\beta$ in Xe.

\subsubsection{Detection in Liquid Xenon}

A comprehensive and clear description of the physics and experimental data on detection of ionizing radiations in noble liquids has been produced by Aprile et al~\cite{Aprile:2006}. Energetic particles in liquid xenon produce observable signals through three processes: one has direct ionization producing free electrons (1), one can excite Xe dimers which decay giving UV light (2), or one can have UV light produced by capture of an electron again producing a dimer which decays giving light (3). To fully measure the energy of the primary particle, one must measure each of these modes. In detectors such as EXO, wire grids or pads are used to measure mode 1 while light sensors (LAAPDs or SiPMs) determine the sum of modes 2 and 3. This greatly improves the energy resolution compared with a straight ion chamber signal but to obtain the best resolution, one needs to separate modes 2 and 3 as they will contribute differently to the total energy sum. Indeed one expects mode 3 to have a similar energy per production as that of mode 1 while mode 2 may be different. In addition, the photon detection efficiency is limited to about 15\% (for the case of SiPMs leading to less than complete energy determination. The measured resolution in EXO was about 1.15\% ($\sigma$/E)~\cite{PhysRevLett.123.161802}. The Fano limit is about 0.2\%~\cite{Doke:1976zz}. These concepts lead to ideas on how the energy resolution might be improved.

\subsubsection{Charge Only, CerenkOv Light Assisted Test Experiment {(CHOCOLATE)} Concept}

The first change I would suggest is to follow the work of Doke and introduce triethylamine (TEA) into the xenon. Doke showed that with a few PPM of TEA he could convert 85\% of the UV photons into electrons by photoionization of the TEA. These can in turn be detected by the electron detectors. There are two advantages of working with charge: the electrons can be detected with close to 100\% efficiency (compared with about 15\% for the photons) and one can determine where the electrons came from. Thus an immediate improvement in resolution would be made due to the factor of 6 increase in the light signal statistics. 

In a TPC, the signal would now look like a concentrated group of about 105 primary electrons within a radius of 1-2~mm surrounded by a halo of about 105 electrons coming from the converted photons at a radius up to a few centimetres (fixed by the TEA concentration). One could measure, for each cluster, both light and charge with good resolution getting an accurate measure of the charge to light ratio. Both the `light' and charge would be measured with the same devices thus largely removing the systematic error introduced by the need to calibrate the efficiencies. However, a new systematic uncertainty would be added due to the uniformity of the TEA distribution. 

One expects more light from a double beta decay cluster because it will have two high charge density clusters at the ends of the two electron tracks and these are the locations where one would have the maximum recombination rates. With gas counters it is well known that one can get a discrimination against electron and gamma backgrounds of a factor of about 20 by observing such clusters and some or all of this may be available to the charge to light ratio measurement.

There are drawbacks to using TEA in the detector. First, one notes that most dual-phase detector modes will be difficult because any electroluminescence photons produced can cause photoionization of the TEA and one gets a runaway positive feedback. Second, as most of the photons have been absorbed one loses the start signal for the event which is needed to locate the event in the Z direction. There may be sufficient Cerenkov radiation below the photoionization threshold to provide this information. There may also be other radiations from the Xe that might help. In gaseous Xe there are as may IR photons as UV and while these are expected to be suppressed in liquid, these may also provide a small boost to the signal.

Thus our first detector will be referred to as the Charge Only, CerenkOv Light Assisted Test Experiment, or CHOCOLATE.

\subsubsection{Pixelated Anode}

The EXO detectors use pads or no-gain wires to detect the anode signals. These detection modes are quite slow ($\mathcal{O}(\mathrm{\mu s})$), have rather poor spatial resolution (several mm), and have rather high electronic noise (several hundred electrons per pad/wire) which obscures important features of the data. For example, in EXO-200 an important background suppression came from selecting events in which only one cluster was observed. The rejection of the ${}^{228}$Th gamma line was a factor of 6 but one would have expected a factor of 40 (the ratio of the Compton and pair-production cross sections to that of the photoelectric process). One expects a further factor of 7 if one could identify low energy clusters and achieve better separation of the closely spaced clusters. 

A lot of work has been taking place recently on pixelated detectors. A recent workshop on such detectors was held in the UK and in a summary talk five examples were quoted with spatial resolutions of $\sim 50~\mathrm{\mu m}$ and with sub-electron noise levels. For some devices, it would seem possible to also get timing resolutions of order nanoseconds. With such a system one could get very clean identification of event clusters as well as some of the lower energy processes. Ideally one would like to get the full charge $Q(x,y,t)$ image but this may be beyond the current state of the art. It does look possible to get $Q(x,y)$ integrated over time, and $Q(t)$ integrated over $(x,y)$, and this would provide everything needed. 

Better spatial resolution suggests other ways in which the detection could be improved. For example, the electron tracks in liquid xenon are the same as those seen in gas but compressed from $\sim$10~cm to about 1.5~mm. Perhaps one could get the same gamma discrimination as seen in gas detectors (a factor of 20) based on two high-ionization regions, from the spatial signal. Even if one could not resolve the details, a measure such as the extent of the charge along the drift direction might allow the columnar contribution to the recombination signal to light to be estimated.

The main limitation to the use of spatial information will come from diffusion of the electron cloud. It may be possible to use deconvolution techniques or artifical intelligence/machine learning to recover the signal. This requires further study to see if it is possible.

\subsubsection{Positive Ion Enabled (PIE) detector}

The diffusion limitation of the previous section suggests that another approach might be useful. Some years ago, Martell et al.~\cite{Martoff:2000wi} pointed out that in most media, but especially in noble gases and liquids, the electrons are very non-thermal and thus have very high diffusion levels. Molecules, on the other hand will remain quite close to thermal. Diffusion is small, even with long drift time because, from the Einstein relations D=$\mu$kT/q, D is small when the mobility is small.

We can build on and adapt this idea to our CHOCOLATE detector. The ionizing event will leave about 105~Xe ions but these will quickly charge transfer with the TEA to form TEA+. These ions can then drift towards the cathode where they would be detected by a pixel detector. This might be a CMOS device or perhaps a charge-enabled CCD device. Such an image could be very sharp indeed. (This concept probably would not work in pure xenon detectors as the mobility of the Xe+ is considerably higher than that of other ions but most of the Xe would charge exchange with impurities at some point in the drift).

There is an obvious drawback to using positive ions. The mobility of ions in liquid xenon is very low and quite high drift voltages would be required. At high fields the electron recombination is supressed, and we may lose the two-electron signature coming from charge-to-light ratio. On the other hand, if we get good imaging of the events, we may not need this. In any event, a lower recombination is probably still an advantage in getting good energy resolution. Another option might be to operate at modest voltage until a candidate event is seen and then raise the drift field. The positive ions will not have gone anywhere!

Finally, if the positive ion signal can provide a modest time signal, the light start signal is no longer required to determine the z coordinate for the event. Also, clearly, if positive ion detection is to be used, we must design the whole detector to be high-voltage capable.

The limitation to resolution of positive ions may be caused by turbulence in the liquid. It should be possible to set up the conditions in the xenon to avoid this. The Reynolds number is low enough to suggest laminar flow and with careful temperature control on the boundary of the Xe this should be achieved. Heat from the pixel detectors at the bottom could be an issue but one could imagine leaving them off until a candidate event is observed in the anode and then only enabling the devices in the region of the expected signal.

\subsubsection{Use an active veto}

Many detectors employ an active veto to supress gamma rays or neutrons that scatter out of the Xe detector. In the EXO detectors, there is an outer region filled with a heat transfer fluid called HFE-7000. This has some useful properties including a freezing point below that of Xe, a high density at low temperatures, low contamination by radioactive impurities (even straight from the container!), and it is very transparent. It might be possible to make it scintillate.

\subsubsection{The CHOCOLATE PIE concept}

Finally we put these ideas together to form the CHOCOLATE PIE detector (Charge Only, CerenkOv Light Assisted Test Experiment with Positive Ion Enabling). A cartoon is shown in Fig.~\ref{fig:chocolate_pie_concept}). Some features of this detector are as follows:

\begin{figure}[h]
    \centering
    \includegraphics[width=\linewidth]{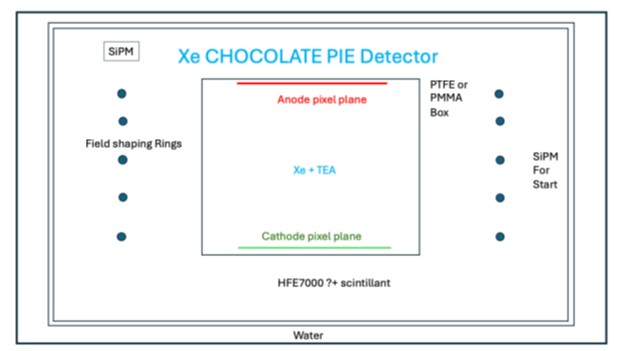}
    \caption{Concept for a Charge Only, Cerenkov Light Assisted Test Experiment with Positive Ion Enabled}
    \label{fig:chocolate_pie_concept}
\end{figure}

\begin{enumerate}
    \item We contain the Xe with an acrylic tank.
    \item We use the HFE-7000 (or potentially some other suitable fluid) buffer to provide all shielding of external gamma rays.
    \item We remove all the voltage gradient electrodes out of the Xe. We make them out of acrylic tubes, perforated to fill with the HFE-7000 and coated with Clevios to be conductive to reduce metals.
    \item All metals (the main sources of background in EXO) are excluded from the Xe. 
    \item The acrylic tank can have corrugated walls to give better high-voltage standoff.
    \item Use spark gaps and/or equivalent approaches to protect the outer region from damage in event of breakdown.
    \item The HFE-7000 (or substitute) is loaded with scintillant.
    \item SiPMs in the HFE are used to detect the Cerenkov radiation and light and to detect the scintillation (wavelength filters will separate these functions).
\end{enumerate}

Some additional features of the detector are:

\begin{enumerate}
    \item Self absorption (``self-shielding'') in the xenon is not required so all of the xenon is fiduicial. 
    \item As a result of the first point, the Xe costs are kept to a minimum.
    \item Also as a result of the first point, the detector can be modular. This might allow the detector to be built and operated in several countries.
    \item Having a modular system allows detectors to come on in a sequence that matches the delivery of Xe.
    \item Having a distributed system probably has real political advantages.
    \item Xe and HFE-7000 are good insulators with dielectric strengths comparable with high pressure $\mathrm{SF_6}$. %
    A 1~tonne detector is about the size of a 3~MV van De Graff generator so $\mathcal{O}(\mathrm{MV})$ voltages should be possible.
\end{enumerate}

\subsubsection{Conclusions}

The objective of the this study was to identify possible ways to improve the energy resolution and background rejection in a liquid xenon detector in order to allow detection of $0\nu\beta\beta$ at in the NO scenario. The main improvement in resolution would come from the 6-fold increase in counting statistics for the light portion of the signal but better control of the recombination vs direct production should also help. For background reduction, the potential improvements include a factor of 7 from better single-site selection, a factor of 20 from identification of two high-ionization densities with a cluster, and a factor of perhaps 5 from removal of all metals from contact with the Xe. The background rejection will be enhanced by the better energy resolution. In nEXO (Section~\ref{sec:nexo}), 80\% of the expected background comes from the 2440~keV line in the decay of ${}^{214}$Bi which is degenerate with $Q_{\beta\beta}$; this overlap will begin to separate with the improved resolution suggested here. The product of these background reductions would be of order 1000 but probably not all will be fully realized. However, the objective of an overall factor of 100 is, perhaps, not unreasonable.

\subsubsection{Appendix 1: Short-term R\&D objectives}

A number of ideas introduced above need to be studied in small scale experiments to test how well reality matches with the ideal. Some of these are as follows:

\begin{enumerate}
    \item A literature study of electron-sensitive pixelated detectors should be made to identify possible candidates for the anode. A small TPC using such a structure should be made operating with liquid xenon.
    \item Measure the response of a liquid Xe TPC loaded with TEA. Measurements should be made as a function of drift field and of the TEA concentration. Use of an external 2.614~MeV gamma source and a gamma counter to tag the event as producing either a Compton-scattered photon or a 511~keV annihilation gamma ray would allow the study of 1- and 2-electron clusters.
    \item A measurement of diffusion of positive ions and especially TEA should be made at various drift fields.
    \item A literature study of pixelated devices sensitive to positive ions and operable in liquid Xe should be made.
    \item An investigation of possible fluids for the veto should be carried out, including a study of scintillating additives.
    \item A Monte Carlo simulation study of the detector should be developed. The study should look at the desirability of modularization of the detector and at the performance vs. isotopic enrichment for the detector. The study should include the possibility of doing dark matter searches with this detector.
\end{enumerate}

\subsubsection{Appendix 2: Possible timelines}

Estimating R\&D timescales is notoriously difficult but if there is sufficient interest, the above research objectives might be achieved in a four-year time frame. At that point it will be clear whether a case can be developed for a major venture towards the NO scenario. At this point it would be appropriate to start work on a prototype to be located underground to demonstrate that all performance objectives, including the background rejections and energy resolutions, can be achieved on a large scale (perhaps 1~tonne). By this time it would be important to have developed an international partnership to carry out the venture. This would also be the time to develop a strategy for procurement of xenon with a long term commitment that would encourage industries to invest in further extraction facilities and probably enrichment facilities. At this point significant lab resources would be required. Allowing 3-4 years to design and construct the prototype together with a year or two of operating and analysis, one can anticipate a proposal for the construction of the first module in about year 10. First data might start to flow in year 15.

Some short-term milestones:

\begin{enumerate}
    \item Assess the interest in the Canadian collaboration in participation - mid summer 2025.
    \item Possible letter of intent to NSERC for R\&D support Aug. 1, 2025.
    \item Possible sidelines meeting during McDonald Institute national meeting in Ottawa (early August, 2025).
    \item Decide on R\&D topics to be explored in Canada.
    \item Start to build international participation at TAUP in Aug. 2025.
    \item Start work toward a formal collaboration at the end of 2025.
\end{enumerate}

\subsection{Beyond SNO+}
\label{sec:beyond_snoplus}

\authorline{Author: M.~Chen}

\bigskip

I'll briefly outline some considerations for how future $0\nu\beta\beta$ experiments should approach the design of future experiments (with normal mass ordering sensitivity). An experiment's sensitivity is determined by the quantity of double beta-decaying isotope that can be deployed and counted, efficiently, in a detector, compared to the amount of background in the detector that produces similar signals. It's as simple as evaluating signal over background. When thinking about a future experiment, it is {\bf essential that scaling up quantities of the isotope be both logistically feasible and affordable}. Tellurium satisfies this top-level requirement. It is important that a scalable experimental technique for $^{130}$Te be developed, otherwise the field will reach (if it hasn't already) limitations due to economics and logistical feasibility driven by isotope procurement alone.

On the subject of backgrounds, as experiments increase quantities of isotope, there is an increasing need for background control. Background counts appear in the region of interest (ROI) and are often quantified as a \textit{Background Index}: counts/keV/kg/yr, which equals the rate per width in the ROI, per unit mass of deployed isotope. There are five ways (generally speaking) that a future experiment can improve backgrounds:
\begin{enumerate}
    \item\label{chen_resolution} An experiment could {\bf improve its energy resolution}, making the ROI and ``per keV'' smaller.

    \item\label{chen_radiopurity} An experiment could improve its background counting rate by {\bf improving radiopurity and/or shielding}. This includes being cognizant of potential cosmogenic backgrounds that could be problematic at shallow depths underground. Purifying materials underground, to reduce cosmogenics and prevent their reactivation, is a strategy that SNO+ has adopted.

    \item\label{chen_bkgtagging} An experiment can also reduce backgrounds by employing {\bf background tags/ID}, enabling backgrounds to be rejected event-by-event, or statistically constrained.

    \item\label{chen_signal} The double beta decay signal consists of two electrons (the recoiling nucleus is insignificant); {\bf event topology} reconstruction could help to distinguish this signal from background. This is a specific form of background tagging/ID.

    \item\label{chen_sigtagging} Lastly, an experiment could attempt to {\bf tag the daughter nucleus}, thereby eliminating most potential backgrounds apart from $2\nu\beta\beta$ decays which produce the same daughter nucleus.
\end{enumerate}
These five approaches to background reduction make the complete set. $0\nu\beta\beta$ produces two electrons (\ref{chen_signal}) with summed energy equal to the decay endpoint (\ref{chen_resolution}), and a daughter nucleus (\ref{chen_sigtagging}). Backgrounds must either be reduced intrinsically (\ref{chen_radiopurity}) or by identifying ways in data analysis to distinguish them from signals in the detector (\ref{chen_bkgtagging}).

Now, considering $^{130}$Te, we can think about each of these points for background reduction. For point~\ref{chen_resolution}, the approach using Te-loaded liquid scintillator suffers from poor energy resolution. Once can view this as a \textit{target of opportunity} for future efforts at improvement, where there is space for people to make breakthroughs or discoveries. For example, as proposed in my comment during the SNOLAB Future Projects Workshop, what if we could figure out a way to deploy Te in liquid argon? Liquid argon scintillation has four times the light yield of organic liquid scintillator. There may be other ways to improve energy resolution (including returning to CUORE's approach with Te crystals in a cryogenic bolometer) for an experiment with tellurium.

For point~\ref{chen_radiopurity}, all experiments are working on improving radiopurity and shielding. This is an aspect that the Canadian astroparticle physics community at SNOLAB excels at and can continue to lead. SNO+ will be pioneering the purification of tellurium in the underground plant that has been built and is undergoing commissioning at SNOLAB\@. This work is just starting and, as in many experimental endeavours, there may be future improvements that will arise as we learn more by doing.

For point~\ref{chen_bkgtagging}, again, all experiments are pursuing many background tagging strategies to differentiate backgrounds from the signal. An example is single-site/multi-site background discrimination. What isn't widely appreciated is that the loaded liquid scintillator approach has strong capabilities here. SNO+ has, for example, developed single-site/multi-site event classifiers. Gamma-ray backgrounds from external sources or from potential cosmogenic backgrounds can be identified and statistically constrained in this manner. There is the possibility of exploiting Cherenkov light to provide directional information, thus rejecting backgrounds from $^{8}$B solar neutrinos. The classic $^{214}$Bi-Po delayed coincidence provides strong suppression of radon backgrounds in the $^{130}$Te ROI\@.

For points~\ref{chen_signal} and \ref{chen_sigtagging}, it is exciting that the xenon double beta decay experiments are incorporating these strategies, most notably NEXT (Section~\ref{sec:topic_next}) with its gas TPC detector that resolves two electron tracks, and both NEXT and nEXO (Section~\ref{sec:nexo}) doing R\&D on barium tagging. What about for $^{130}$Te? The double beta decay of Te has Xe as the daughter isotope. Nobody has come up with ideas for how to do xenon tagging yet, but that doesn't mean it isn't possible. Here we find another potential target of opportunity. What can be asserted (thinking very generally here) is that solid detectors would probably have a much harder time achieving daughter isotope tagging than gas or liquid detectors, and thus doped liquid scintillator remains a viable medium for such fanciful considerations.

\paragraph{Conclusions}
\begin{enumerate}
    \item Developing a scalable experimental technique with $^{130}$Te is important for the future of the $0\nu\beta\beta$ field as it seeks normal mass ordering sensitivity and can be achieved in a practical way.
    \item Ways that future experiments can improve backgrounds: detector resolution, radiopurity and shielding, background tagging / particle ID (including event topology reconstrction), and tagging of the daughter nucleus.
    \item Specific approaches for such a future experiments include: deploying Te in liquid argon, purifying Te in underground plants, exploiting Cherenkov light to provide directional information, improving single-site/multi-site event classifiers, and tagging the Xe daughter of Te double beta decay.
\end{enumerate}%
\subsection{Pursuing the Cosmic Neutrino Background}

\label{sec:topic_CnuB}

\authorline{Author: A.~Arvanitaki}

\bigskip

The cosmic neutrino background is a relic of the Big Bang, and if discovered, would provide us with a snapshot of the Universe from a time when it was just a fraction of a second old. Unfortunately, these relic neutrinos are non-relativistic, meaning they carry very little energy, and their interactions have negligible impact. Even in the case of coherent elastic scattering, where the corresponding interaction rates can be sizeable, scaling quadratically with the number of target constituents, i.e., $N^2$. 

Aiming at improving the possibility of detection of these relic neutrinos, we recently set out to explore new processes that could further enhance energy transfer during the interaction~\cite{Arvanitaki:2024taq}. We found that when an ensemble of two-level systems (e.g., nuclear or electron spins in a magnetic field, or any two levels in an atom or nucleus) is prepared in a coherent atomic state, where each atom is in a superposition of ground and excited states, inelastic scattering or absorption processes can also exhibit interaction rates that scale as $N^2$. We called these superradiant interactions, as they generalize Dicke superradiance exhibited by atoms emitting or absorbing photons~\cite{Dicke:1954zz}. Classically, such a process is akin to coherent scattering from a classical time-dependent potential, which can introduce sidebands in the energy of the scattered particle. Unlike traditional coherent elastic scattering, these processes alter the internal state of the target system, transferring more energy than that delivered to the center of mass. Detecting this change in the internal state of the target requires taking its quantum properties into account; we find that there are observables that go beyond net energy transfer and are sensitive to the sum of excitation and de-excitation rates. Such processes are relevant for any cosmic relic. In the case of the cosmic neutrino background, we find that for a 10~cm polarized nuclear spin ensemble, these relic neutrinos can produce interaction rates of $\mathcal{O}$ (1~Hz) -- an enhancement of over $10^{21}$ compared to incoherent rates. Similarly, the QCD axion or dark photon DM may be absorbed or emitted by much smaller systems at similar rates and could be detected using nuclear spin systems designed to exploit macroscopic coherence.

\paragraph{Summary}

Superradiant interactions offer a new class of observables, providing a pathway to ultra-low-threshold, quantum-enhanced detectors for relic particles. These findings are part of a broader shift in experimental particle physics, where quantum-enhanced metrology is emerging as a key frontier. SNOLAB is one of only a few laboratory facilities in the world singularly focused on the study of elementary particles such as neutrinos and beyond, hosting groundbreaking experiments in the search for DM. In this context, the CUTE facility provides a unique opportunity as a test bed --- or a temporary host --- for such fundamental physics experiments, offering a cryogenic R\&D environment often necessary for quantum-sensing setups. Given that the size of these setups is much smaller than that of traditional particle physics detectors, SNOLAB is well-positioned to not only be a permanent home for fundamental physics experiments, but to also expand on the vision behind CUTE and become an incubator for quantum sensing experiments probing DM, relic neutrinos, and other exotic phenomena with unprecedented sensitivity.%

\newpage
\section{Conclusions: Summary of Requested Capabilities}

\authorline{Co-Authors: D.~M.~Asner, M.~Diamond, S.~J.~Sekula}

\bigskip

Below is a collection of the additional capabilities or facilities, implied or suggested by the authors of this community report, that would be useful or necessary for future projects hosted at SNOLAB.

\begin{itemize}
    \item Upgrade of the biological laboratory facility to meet the Level-III biohazard safety standard, and in parallel, consideration of what is required to host significantly more complex animal models (see Section~\ref{sec:topic_Deep Underground Biology}).
    \item Cryogenic distillation column of sufficient scale and throughput to be of use in both argon and xenon programs which may choose to site themselves at SNOLAB (e.g. ~\ref{sec:argolite}). Installation in Shaft 8 is suggested.
    \item Radon-reduced air availability in the Cube Hall (and potentially additional spaces) to prevent increased surface background levels due to absorptions or adsorption of radon and implantation of radon daughters (see Sections~\ref{sec:gadmc_summary}). A reduced-radon space in the surface clean lab facility is likely also necessary to support activities there. 
    \item Upgrades to the existing chemistry facility, including a dedicated setup for radon removal and noble liquid target purification (e.g. Section ~\ref{sec:gadmc_summary}).
    \item Large-scale cryogenic storage and recovery systems, for argon and/or xenon (see Sections~\ref{sec:gadmc_summary}, \ref{sec:nexo}, \ref{sec:topic_next}).
    \item Expansion of personnel that are likely to be required by large, new projects. This would include additional engineers, scientists, and project management staff, at a minimum. Some entirely new skills may required, e.g. in the case of the NEXT program (see Section~\ref{sec:topic_next}) where one or more experts in high pressure gas are noted. In addition, some small experiments might require mK-cryogenic platforms and low-vibration systems, necessitating additional skilled people development in these areas (e.g. Sections~\ref{sec:helios},~\ref{sec:SmallScaleDM},~\ref{sec:cryo_solid_detectors}).
    Larger experiments, or an expansion of the number of experiments, would likely require a compensating increase in the material screening personnel.
    \item More large water tanks outside the ladder lab area, for experiments that are currently ladder-lab-scale but grow in size in the future (e.g. Section~\ref{sec:sbc_summary}).
    \item The expected pressure from a range of projects is expected to require a general expansion of underground laboratory space.  The aggregate of the space anticipated in this report for new experiments, new utilities, and new underground capabilities will exceed the current underground capacity. This would motivate an expansion including additional drifts and ladder labs. In addition, an entirely new cavern space in SNOLAB would be required to host a project of future-generation, kilotonne-scale capabilities (see Section~\ref{sec:theia}).
    \item Capability and capacity to advance chemical and physical processes potentially needed for normal-ordering sensitivity in the $0\nu\beta\beta$ space. For example, doping of scintillators (e.g., argon) with high-abundance double-beta-decay-capable isotope (e.g., tellurium), or tagging of xenon atoms produced by the decay of tellurium (see Section~\ref{sec:beyond_snoplus}).
    \item Preparation for increased demand on the CUTE facility, in the case that more and more new technologies desire a place to test their designs in a well-shielded and cryogenic environment (e.g. Sections \ref{sec:cryo_solid_detectors},~\ref{sec:globolo},~\ref{sec:topic_CnuB}). These tests will very likely also lead to scientific output, such as experimental constraints on nuclear matrix element predictions. In addition, the activities at the CUTE facility involving qubits, if successful, may lead to the need to integrate qubits alongside new detector targets with the qubits serving as the active detector elements (see Section~\ref{sec:smallscaleDM_summary}). In general, there is a sense of increasing pressure on the underground CUTE facility from anticipated projects, meriting the consideration of an \textit{additional} CUTE facilty underground. This could be placed in an existing water tank in the current ladder labs or be part of a new ladder labs if the laboratory were expanded.
    \item Construction of facilities for underground crystal growth, crystal fabrication, and metal manufacturing. These would avoid the backgrounds from cosmogenic tritium production that are problematic in some pieces of apparatus  (e.g. Section~\ref{sec:cryoSD_summary}), and would facilitate low-mass DM detectors using novel detector materials (see Section~\ref{sec:SmallScaleDM}). Quality assurance will likely include a range of new capabilities for SNOLAB, including advanced microscopy (e.g., scanning tunnelling microscopes or similar) and expanded chemical quality assurance and assay capabilities (e.g., more ICP-MS throughput as through additional instrumentation). Increased manufacturing capability is likely to necessitate expanded surface assembly areas to stage larger components, assembled at site, before they move underground.
\end{itemize}
\newpage
\section{Acknowledgements}

SNOLAB is located on the traditional territory of the Robinson-Huron Treaty of 1850, shared by the Indigenous people of the surrounding Atikameksheng Anishnawbek First Nation as part of the larger Anishinabek Nation. We acknowledge those who came before us and honour those who are the caretakers of this land and the waters.

M.~Diamond and Z.~Hong would like to acknowledge the support of the Arthur B. McDonald Canadian Astroparticle Physics Research Institute and of NSERC, discussions with Aviv Noah (HVeV project), discussions with Taylor Aralis, Yoni Kahn, Noah Kurinsky and Jamie Ryan (SPLENDOR project), and discussions with Enectali Figueroa-Feliciano (HONEYCOMB project).

The work of M. R. Lapointe and collaborators is funded through a Natural Sciences and Engineering Research Council of Canada (NSERC) Alliance Grant in partnership with the Nuclear Innovation Institute (NII).

The work of Y. Kahn and collaborators is supported in part by DOE grant by DOE grant DE-SC0015655.

K.~J.~Vetter and L.~A.~Winslow would like to acknowledge the support of the CUPID collaboration and discussions with the ACCESS project. Their work is supported by the United States National Science Foundation grant 2411650 and by the United States Department of Energy grant DE-SC0011091.

The Modane Underground Laboratory (LSM) would like to thank its staff for support through underground space, logistical and technical services. LSM operations are supported by the CNRS, with underground access facilitated by the Société Française du Tunnel Routier du Fréjus.

The China Jinping Underground Laboratory (CJPL) would like to thank its staff for providing underground space, logistical and engineering services. CJPL is jointly operated by Tsinghua University and Yalong River Hydropower Development Company.

We are grateful to Pietro Giampa, whose recommended process for generating and writing this report was followed with great success.

\newpage

\bibliographystyle{elsarticle-num}
\bibliography{main}

\end{document}